\documentclass[aps,twocolumn,prc,nofootinbib,a4paper]{revtex4}
\usepackage{graphicx}
\usepackage{amssymb}
\usepackage{bm}

\begin{document}
\title{Description of superdeformed bands in light $N=Z$ nuclei
by cranked HFB method}
\author{Makito Oi}
\affiliation{Department of Physics, University of Surrey, Guildford, GU2 7XH, 
  Surrey, United Kingdom}
\email{m.oi@surrey.ac.uk}
\date{August 17, 2007; Ver.1.2}

\begin{abstract}
Superdeformed states in light $N=Z$ nuclei are studied
by means of the self-consistent cranking calculation 
(i.e., the P$+$QQ model based on the cranked Hartree-Fock-Bogoliubov method).
Analyses are given for two typical cases of superdeformed bands 
in the $A \simeq 40$ mass region, that is,
bands where backbending is absent ($^{40}$Ca) and present ($^{36}$Ar).
Investigations are carried out, particularly for the following points:
cross-shell excitations in the sd and pf shells; the role of
the g$_{9/2}$ and d$_{5/2}$ orbitals; the effect of the nuclear pairing; 
and the interplay between triaxiality and band termination.
\end{abstract}

\maketitle

\section{Introduction}

New ``islands'' of superdeformation (SD) were found,
after nearly fifteen years of systematic search,
in the nuclear chart around the $A\simeq 40$ mass region 
(e.g., $^{36}$Ar \cite{SMJ00} and $^{40}$Ca \cite{ISR01}).
Surprisingly, these light and symmetric (i.e., $N=Z$) nuclei 
in the latest SD ``archipelago'' are  magic and near-magic systems,
whose ground states have spherical shape.
These light nuclear systems with magic and near-magic numbers 
need cross-shell excitations, involving both the sd and pf shells,
in order to produce collective degrees of freedom necessary 
for the formation of SD states.
The corresponding shell-model space becomes inevitably very large.
However, modern high-performance computation systems are
quickly advancing to allow shell-model diagonalization to be executed 
if the minimum and reasonable truncations are justified in the model space.

Mean-field descriptions have a numerical advantage in reducing 
the dimension over exact diagonalization, 
owing to the ansatz for a many-body wave function 
(for example, the Slater determinant in the Hartree-Fock theory). 
The mean-field approach was exclusively applied \cite{BHR03} 
to the early studies of SD states known before 2000 \cite{SFC97}, 
such as in the $A\simeq 80$  (e.g., $^{84}$Zr$_{44}$ \cite{JBB95}),
$A\simeq 150$ (e.g., $^{152}$Dy$_{86}$ \cite{TNN86}), and
$A\simeq 190$ (e.g., $^{194}$Hg$_{114}$ \cite{THW90}) mass regions.
The main reason is that nuclei in these SD archipelagoes
belong to heavy- and medium-weight classes. They are still out of reach of
the shell-model diagonalization approach using the full model space
in the relevant valence shells. 
The mean-field method has other advantages, particularly related to
intuitive understanding of many-body systems, 
such as nuclear deformation and nuclear superconductivity.

At present, it is true to say that
neither mean-field calculations nor truncated shell-model diagonalizations
are dominantly superior to their counterpart.
They are complementary at the moment.
Many theoretical studies using these two approaches followed
after the experimental reports were published on $^{36}$Ar and $^{40}$Ca.

Shell-model diagonalizations were performed by Caurier et al. 
\cite{SMJ00,CNP05,CMN07} and Poves \cite{Po04}, 
with the removal of the d$_{5/2}$
orbital from the sd-pf shells and with the corresponding effective interaction.
Recently, they attempted to give a consistent description of the SD and 
normal deformed states in $^{40}$Ca \cite{CMN07}.
These calculations with the truncation were very successful 
in reproducing the experimental energy spectra.

The first attempt through the mean-field approach
was carried out with the cranked Nilsson model in the
original paper \cite{SMJ00}. In the $^{40}$Ca paper \cite{ISR01}, the
cranked RMF (relativistic mean field) method was applied. Both of the methods
ignored the pairing correlation, so that the energy spectra of 
the low-spin members ($J\alt 10\hbar$) in the SD bands were not well reproduced.

Long and Sun then applied the projected shell model (PSM). In this model,
basis are produced through angular momentum projection 
onto the Hartree-Fock-Bogoliubov (HFB) states,
obtained with the P$+$Q$\cdot$Q two-body interaction \cite{LS01}.
A merit in this framework is that the pairing correlation 
is properly treated. As a consequence, 
a better agreement was obtained with the experimental data.
However, the calculation was restricted 
to an axial symmetric shape (with the deformation parameter fixed 
all the way from low- to high-spin regions),
so that the shape evolution of the system, particularly
the triaxial degree of freedom in response to the Coriolis force,
cannot be discussed in this model. 

Variable deformation is an important 
degree of freedom in a rapidly rotating nucleus.
For example, the band-termination phenomenon for the SD band
is predicted by the cranked Nilsson model, which gives
a continuous evolution in triaxiality  towards 
the non-collective oblate deformation ($\gamma=-60^{\circ}$) in the band limit
\footnote{The sign convention for $\gamma$ in this study is opposite to the
so-called Lund convention.}.
Inakura et al., applied the cranked Skyrme Hartree-Fock (HF) method, 
which does not restrict nuclear shape unlike the PSM, 
but the pairing correlation  was ignored \cite{IMY02}.
Bender, Flocard, and Heenen analyzed the SD bands in the $A\simeq 40$ region
by means of the most sophisticated method, the generator coordinate 
method (GCM) with the projected Skyrme HF+BCS states \cite{BFH03}. 
The Lipkin-Nogami method and particle number projection 
were applied, so that the pairing was properly treated. 
Although nuclear shape can vary through the constraint on
the quadrupole moment, only axial deformation was assumed in the calculation.
In addition, the analyses were restricted only to low-spin states 
($J\le 6\hbar$).
This is because the states with non-zero angular momentum were only 
kinematically created through angular momentum projection (without cranking). 
A dynamical effect originating from shape coexistence 
was considered through the GCM, but
the method underestimated the more important dynamical effect
coming from the Coriolis force,
which plays a major role at high spin.

The aim of this paper is thus to test another mean field approach,
which can handle the pairing correlation, the Coriolis force, and
the evolution of nuclear shape (in particular, triaxiality)
 in a fully self-consistent manner, for the full sd-pf model space.
For this purpose, 
the SD bands in $^{36}$Ar (a case with backbending) 
and $^{40}$Ca (a case without backbending) are analyzed with 
the P$+$Q$\cdot$Q model based on the HFB method \cite{HO95},
in this work.

\section{The P$+$Q$\cdot$Q model of Self-consistent Cranking calculation}

In the current framework, the Hamiltonian contains two terms,
\begin{equation}
  \hat{H}=\hat{H}_0 + \hat{V}.
\end{equation}

The first term $\hat{H}_0$ represents the one-body term and
it is the spherical Nilsson Hamiltonian.
In the second quantization notation, it is expressed as
\begin{equation}
  \hat{H}_0 =\sum_me_ma^{\dag}_ma_m,
\end{equation}
where a pair of operators $(a^{\dag}_m,a_m)$ denotes fermionic operators of
creation and annihilation.
This part is solved exactly
\begin{equation}
  \hat{H}_0|\psi_m\rangle = e_m|\psi_m\rangle,
\end{equation}
and the eigenstates $\{|\psi_m\rangle\equiv a^{\dag}_m|0\rangle\}$ 
(the spherical Nilsson states)
are used as the basis in the following stages.
Index $m$  collectively denotes the quantum numbers 
in the Nilsson model, that is, $(nlj\Omega)$, as well as isospin and parity. 
A time-reversal state of $m$ is denoted as $\bar{m}$.
We use the notation ``$m>0$'' which means $(nlj;\Omega > 0)$. 
In this case, its time-reversal state $\bar{m}$ corresponds to $(nlj;-\Omega)$.
The so-called Nilsson parameters for the spin-orbit and orbit-orbit forces
(denoted as $\kappa$ and $\mu$ in the standard notation) 
are taken from Refs. \cite{NR95} and \cite{BR85}.

The second term $\hat{V}$ represents the two-body part,
and it is the P$+$Q$\cdot$Q interaction in this study,
\begin{equation}
  \hat{V} = -\frac{1}{2}\chi\sum_{\mu=-2}^{2}\hat{Q}_{\mu}^{\dag}\hat{Q}_{\mu}
  -\sum_{\tau=\text{p,n}}G_{\tau}\hat{P}^{\dag}_{\tau}\hat{P}_{\tau},
\end{equation}
where the first and second terms correspond to the particle-hole
and particle-particle channels of the two-body interaction, respectively.
The former interaction is responsible for the long-range correlation
to describe nuclear deformation, while the latter is for the short-range
correlation to handle the nuclear pairing. 
The quadrupole operator and the monopole pairing operator are 
respectively given as
\begin{eqnarray}
  \hat{Q}_{\mu} &=& \sum_{mn}\left(Q_{\mu}\right)_{mn}a^{\dag}_ma_n,\\
  \label{Qph}
  \hat{P}_{\tau} &=& \sum_{m(\in \tau)>0}a_{\bar{m}}a_m.
  \label{Ppp}
\end{eqnarray}

The Hamiltonian is ``diagonalized'' with the basis $\psi_m$,
by means of the mean field approximation.
It corresponds to a procedure to extract
one-body ingredients, $\hat{V}_{\text{MF}}$, from the two-body 
interaction, $\hat{V}$, so as to diagonalize $\hat{V}_{\text{MF}}$.
 The residual part, $\hat{V}_{\text{R}}=  \hat{V}-\hat{V}_{\text{MF}} $, 
is therefore neglected in the approximation.

Remembering that we take the pairing correlation into account,
the HFB ansatz is employed in the variational calculations.
That is,
\begin{equation}
  |\text{HFB}\rangle = \prod_p\beta_p|0\rangle.
\end{equation}
The quasiparticle annihilation-creation operators 
($\beta_p$,$\beta_p^{\dag}$) are related to the original 
annihilation-creation operators ($a_m,a^{\dag}_m$)
through the Bogoliubov transformation,
\begin{equation}
  \left(\begin{array}{c}\beta \\ \beta^{\dag}\end{array}\right)
  =\left(\begin{array}{cc} U & V^* \\ V & U^*\end{array}\right)
  \left(\begin{array}{c}a \\ a^{\dag}\end{array}\right).
\end{equation}
Matrix elements $U$ and $V$ in the Bogoliubov transformation 
correspond to the variational parameters in the HFB theory.
The density matrix $\rho$ and the pairing tensor $\kappa$ are introduced here,
as
\begin{eqnarray}
  \rho_{mn} &=& \langle\text{HFB}|a^{\dag}_na_m|\text{HFB}\rangle 
  = (V^*V^{T})_{mn},\\
  \kappa_{mn} &=& \langle\text{HFB}|a_na_m|\text{HFB}\rangle
  = (V^*U^{T})_{mn}.
\end{eqnarray}
The mean-field approximation of the Hamiltonian thus reads
\begin{eqnarray}
  \label{MF_Hamiltonian}
  \hat{H}_{\text{HFB}} &=& E_{\text{HFB}}+\sum_{p}E_p\beta^{\dag}_p\beta_p\\
  \nonumber
  &=& E_{\text{HFB}}+\sum_{mn}h_{mn}a^{\dag}_ma_n 
  +\sum_{\tau=\text{p,n}}\sum_{mn\in\tau}\Delta_{mn}^{\tau}a_ma_n + \text{h.c.}.
\end{eqnarray}

The one-body component in the particle-hole (ph) channel, represented as $h$,
has the following form.
\begin{eqnarray}
  h &=& e + \Gamma 
  \label{single-hamiltonian}
\end{eqnarray}
The first term $(e)_{ij}=\delta_{ij}e_i$ represents the spherical Nilsson energy.
The second term $\Gamma$ is called the self-consistent potential,
\begin{equation}
  \Gamma_{mn}=\sum_{\mu=-2}^{2}q_{\mu}\left(Q_{\mu}\right)_{mn},
\end{equation}
where the self-consistent coefficient $q_{\mu}$ is given as
\begin{equation}
  q_{\mu}=-\chi\text{Tr}\left(\rho Q_{\mu}\right).
  \label{QQc}
\end{equation}
The coupling constant $\chi$ is determined in the standard manner 
by comparing to the axially deformed Nilsson model 
with deformation $\beta^{\text{0}}$ in the beginning of the variational
calculation.

The one-body component in the particle-particle (pp) channel, 
denoted as $\Delta$ in Eq.(\ref{MF_Hamiltonian}),
describes the pairing correlation. It has the form,
\begin{equation}
  \Delta_{mn}^{\tau} = \frac{1}{2} p_{\tau}^*\left(P_{\tau}\right)_{mn}
  \quad (\tau=\text{p,n}),
\end{equation}
where the pairing matrix element $\left(P_{\tau}\right)_{mn}$ 
is determined from Eq.(\ref{Ppp}) 
and the self-consistent pairing coefficient $p_{\tau}$ is expressed as 
\begin{equation}
  p_{\tau} = - G_{\tau}\sum_{m(\in\tau)>0}\kappa_{m\bar{m}}.
\end{equation}
The pairing-gap energy ($\bar{\Delta}$) is defined as the average
of the matrix elements of $\Delta$, that is,
\begin{equation}
  \bar{\Delta}^{\tau}\equiv \frac{1}{M'}\sum_{m(\in\tau)>0}^{M'}\Delta_{m\bar{m}}^{\tau},
\end{equation}
where $M'=M/2$ is the half the dimension of a subspace 
of given isospin ($\tau$).
In the case of the present separable interaction, the expression for the
pairing gap is simply given as
\begin{equation}
  \bar{\Delta}^{\tau}=p_{\tau}.
\end{equation}
The pairing strength $G_{\tau}$ is determined in the standard manner
by using the Nilsson+BCS calculation with the initial values 
for the pairing gaps ($\Delta^{\text{0}}$) together with the 
$\beta^{\text{0}}$. 

High-spin states are produced with the self-consistent cranking model.
That is, the variational equation,
\begin{equation}
  \delta\langle\text{HFB}|\hat{H}-\omega\hat{J}_x-\sum_{\tau=\text{p,n}}
  \lambda_{\tau}\hat{N}_{\tau}|\text{HFB}\rangle=0,
\end{equation}
is self-consistently solved by means of the gradient method 
under the following two constraints:
\begin{equation}
  \langle\text{HFB}|\hat{J}_x|\text{HFB}\rangle =\text{Tr}\left(\rho j_x\right)
  = J,
  \label{const-J}
\end{equation}
where $J$ is the total angular momentum, and
\begin{equation}
    \langle\text{HFB}|\hat{N}|\text{HFB}\rangle 
    =\text{Tr}\left(\rho\right)= N,
    \label{const-N}
\end{equation}
where $N$ is the total particle number.

In this study, the usual one-dimensional cranking
model is implemented, so that only one component 
of the total angular momentum vector is constrained. 
Quantities $\omega$ and $\lambda_{\tau}$ in the variational equation
are the Lagrange multipliers. The first multiplier is interpreted 
as the rotational frequency, while the second multiplier stands for 
the chemical potential. The presence of the chemical potential is 
due to the introduction  of the BCS-type pairing correlation, 
which breaks the particle number conservation. 
As a result, the mean particle number needs to be constrained  
in the calculation.

The HFB energy $E_{\text{HFB}}$ can be thus written as
\begin{equation}
  E_{\text{HFB}}^J = \langle\text{HFB}(J)|\hat{H}|\text{HFB}(J)\rangle 
  = \text{Tr}\left(\rho h\right)
  - \sum_{\tau=\text{p,n}}\bar{\Delta}_{\tau}^2,
\end{equation}
and this corresponds to the yrast spectrum.

For the model space,
two major shells ($N=2,3$, or the so-called sd-pf shell) 
each for protons and neutrons are used, that is, 
d$_{5/2}$,s$_{1/2}$,d$_{3/2}$ ($N=2$);
f$_{7/2}$,p$_{3/2}$,f$_{5/2}$,p$_{1/2}$ ($N=3$).
This choice is in accordance with the Kumar-Baranger prescription 
for the P$+$Q$\cdot$Q force \cite{KB68}.
When a role of an intruder g$_{9/2}$ ($N=4$) orbital is discussed,
it is also included in the model space.

Further details of the method are available in Ref.\cite{HO95}.

\section{Numerical results}

Two $N=Z$ nuclear systems will be studied in this paper,
which are $^{40}$Ca and $^{36}$Ar.
The former nucleus shows no sign of backbending so far 
(up to $J=16\hbar$),
while the latter has a clear backbending at $J=10\hbar$.
With the P$+$Q$\cdot$Q model based on the cranked HFB approach,
we attempt to describe the superdeformed states in these
nuclei in a self-consistent manner both in the ph- and pp-(hh-) channels.

\subsection{A case of no backbending: $^{40}$Ca}

The SD band of $^{40}$Ca is so far identified up to $J=16\hbar$ \cite{ISR01}.
This rotational band is regular and no backbending is currently observed.

\subsubsection{A role of the d$_{5/2}$ orbital}
As stated above, our model space contains the full sd- and pf-shells.
It is thus possible to examine,
in the framework of the self-consistent mean-field calculation, 
the core-polarization effect, or
the influence coming from the d$_{5/2}$ orbital truncated in the shell-model
calculations.   The occupation numbers 
of each single-particle orbital provides us useful information for this aim.

Before analyzing our own calculations,
it is worth learning the results obtained by others.
In a description of the SD band of $^{40}$Ca 
through the shell-model diagonalization  by Poves \cite{Po04},
the single-particle model space is set to be s$_{1/2}$,d$_{3/2}$ ($N=2$); 
f$_{7/2}$,p$_{3/2}$,f$_{5/2}$,p$_{1/2}$ ($N=3$).
(This choice of the model space was also used for $^{36}$Ar.)
The d$_{5/2}$ orbital is excluded from the valence space 
for the numerical reason.
Within this model space (and the corresponding effective interaction),
the 8p-8h configuration, that is, (s$_{1/2}$d$_{3/2}$)$^{4}$(fp)$^{8}$, 
was proposed for the description of the SD band. 
The calculated result based on this configuration 
reproduces the experimental data quite well with a well-tuned effective
interaction \cite{SMJ00,Po04}.
%
Long and Sun raised the question about the d$_{5/2}$ truncation
 in their paper where they performed
the PSM (projected shell model) calculation\cite{LS01}. 
According to their analysis, the d$_{5/2}$ orbital does not contribute to
the SD state in $^{36}$Ar, but to higher excited rotational bands.
The cranked Nilsson calculation performed for $^{36}$Ar \cite{SMJ00}
is also informative. It was obtained
that only about half a particle in each isospin sector is excited
into higher orbitals from the d$_{5/2}$ orbital. Summarizing these results, 
it can be said that Poves' prescription for the truncation 
might be a  good approximation for a description of the SD band.

\begin{figure*}[htb]
  \begin{tabular}{cc}
    \includegraphics[width=0.35\textwidth,angle=-90]{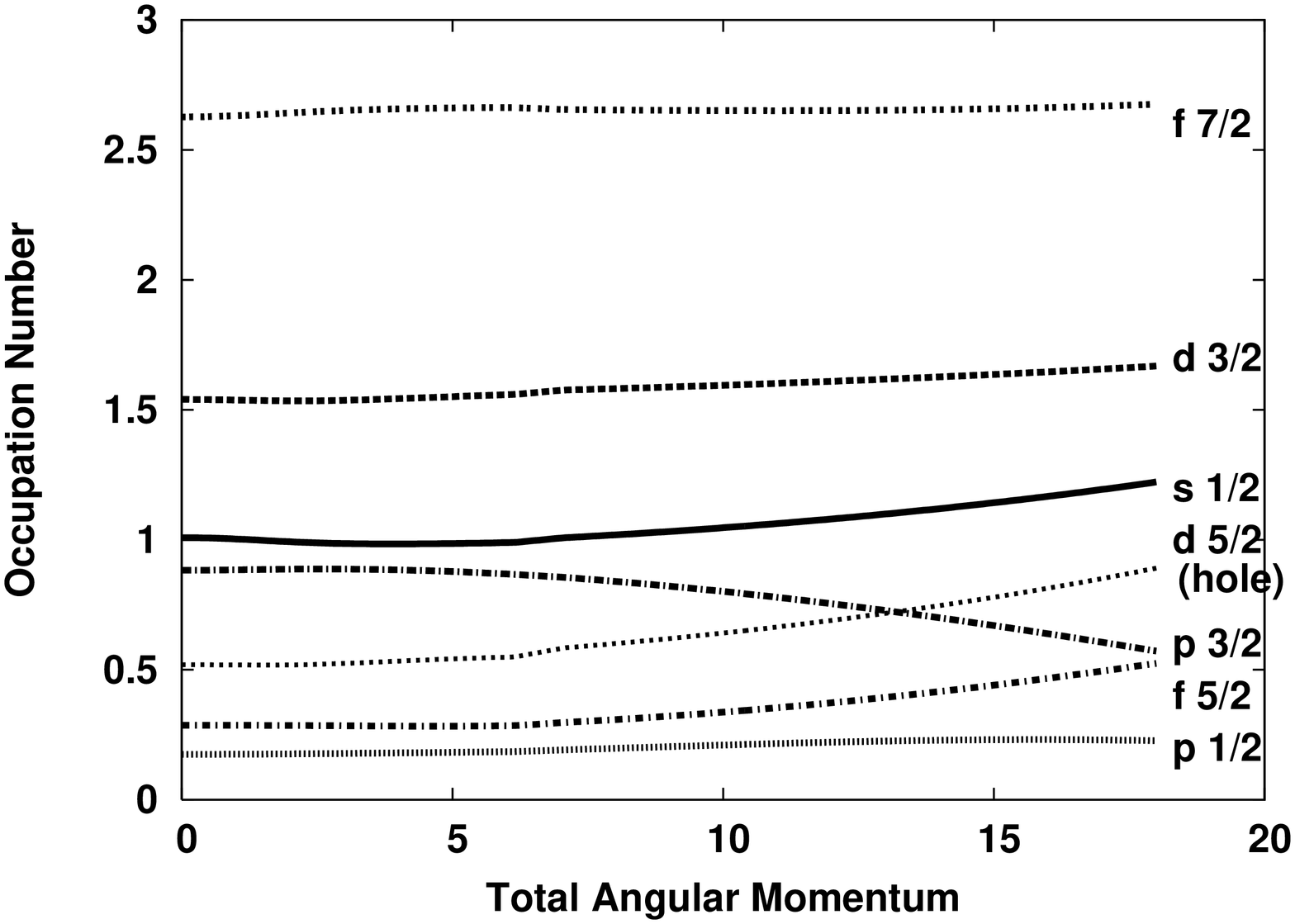}&
    \includegraphics[width=0.35\textwidth,angle=-90]{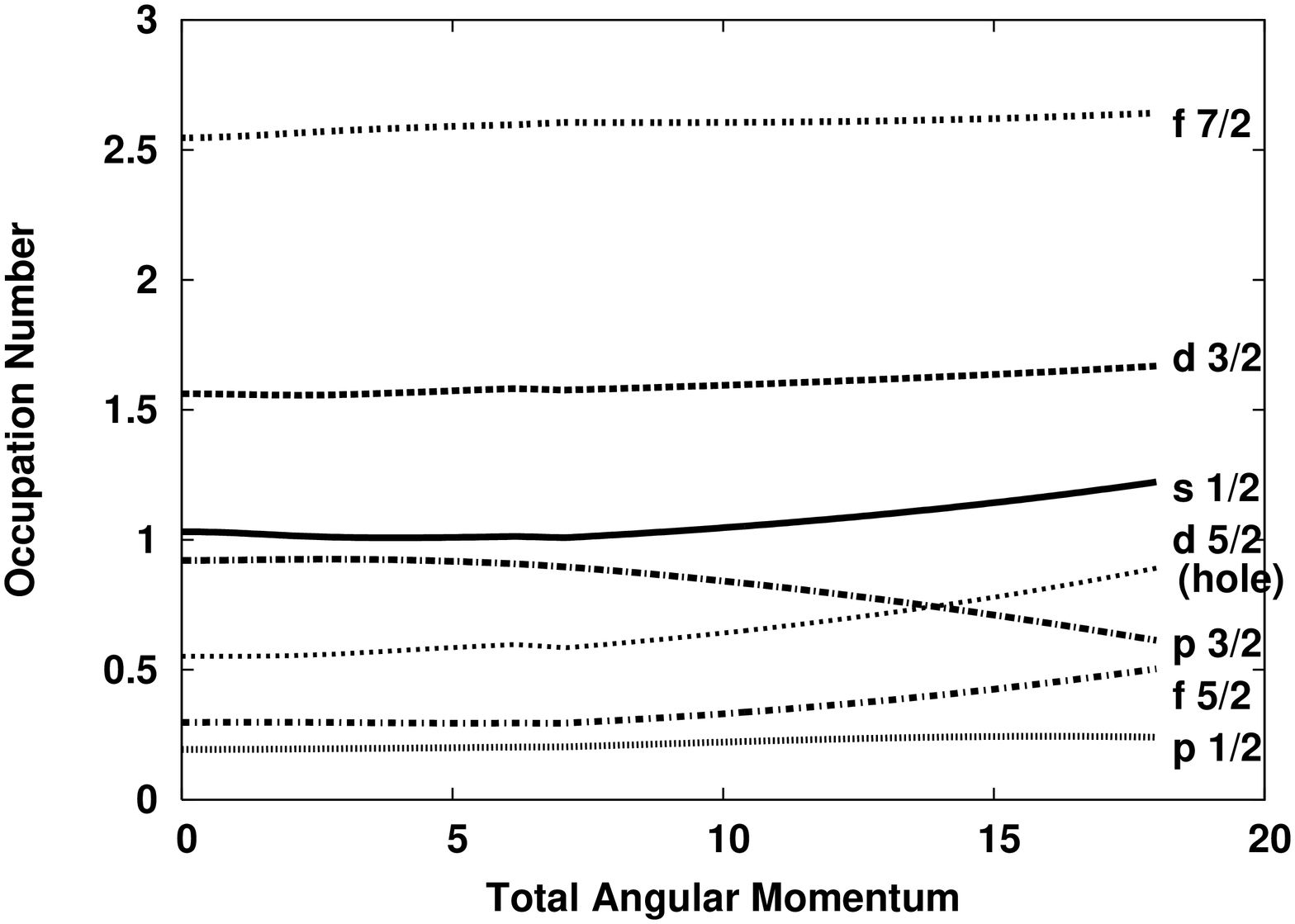}
  \end{tabular}
  \caption{Proton (left) and neutron (right) 
    occupation numbers as a function of the total angular momentum
    for $^{40}$Ca. In this case, the g$_{9/2}$ orbital is excluded from the model
    space.}
  \label{occupation}
\end{figure*}

Now, let us turn to our calculation.
The initial parameters for the self-consistent iterations 
are chosen to be $\beta^0=0.6$ and
$(\Delta^0_{\text{p}},\Delta^0_{\text{n}})=(1.360,1.513)$ MeV.
These initial pairing-gap energies are employed from Ref. \cite{MN95}.

In Fig. \ref{occupation},
the average occupation numbers, which are diagonal elements of the density
matrix, $\rho$,  are presented as a function of the total angular momentum.
The right and left panels in the figure 
show the occupation numbers for protons and neutrons, respectively. 
The graphs look quite similar to each other as a consequence of $N=Z$. 
Following the presentation in Ref. \cite{SMJ00}, the occupation in the d$_{5/2}$
orbital is plotted using the hole occupation number. 
Our result shows (see Table \ref{table1}) 
that only half a particle is missing from the fully 
filled d$_{5/2}$ orbital, which supports the shell-model truncation by Poves.
This result is similar to the cranked Nilsson calculation 
for $^{36}$Ar \cite{SMJ00}. 
However, the number increases at higher spin ($J\agt 16\hbar$)
to reach to 1 (see Table \ref{table2}), 
so that the core polarization may need to be taken into account,
particularly at high spin.

\begin{table}[bth]
  \begin{ruledtabular}
    \begin{tabular}{cccccccc}
      Orbital& d$_{5/2}$ &s$_{1/2}$ &d$_{3/2}$ &f$_{7/2}$ &p$_{3/2}$ &
      f$_{5/2}$ &p$_{1/2}$ \\
      \hline
      Proton & 5.48& 1.01& 1.54& 2.63& 0.88& 0.29& 0.17\\
      Neutron& 5.45& 1.03& 1.56& 2.55& 0.92& 0.30& 0.19\\
      \hline
      Total  &10.93& 2.04& 3.10& 5.18& 1.80& 0.59& 0.36\\
    \end{tabular}
  \end{ruledtabular}
  \caption{Occupation numbers of $^{40}$Ca at $J=0$. 
    The subspace (s$_{1/2}$d$_{3/2}$)
    is occupied by about five ($=5.14$) particles, 
    while the pf-shell is filled with about eight ($=7.93$) particles. 
    The hole occupation number in the d$_{5/2}$ orbital is 1.07 ($=12-10.93$).
  }
  \label{table1}
\end{table}

Table \ref{table1} displays the details of the occupation numbers at $J=0$.
The net occupation number in the d$_{5/2}$ orbitals
are 10.93, that is, about one particle (in the isoscalar basis) 
is excited into upper orbitals, as already mentioned above.
The total numbers of the occupation in the subspace (s$_{1/2}$d$_{3/2}$)
and the pf shell (f$_{7/2}$p$_{3/2}$f$_{5/2}$p$_{1/2}$) 
are about five and eight, respectively. In other words, our calculation
suggests (d$_{5/2})^{-1}$(d$_{3/2}$s$_{1/2}$)$^{5}$(fp)$^{8}$
for the band-head structure of the SD band. To a good extent, 
this configuration is consistent with the 8p-8h structure proposed 
in the shell-model calculation.
 Although our calculation indicates a possible
core polarization, this effect can be minor at low spin.

Table \ref{table2} shows the occupation numbers at $J=18\hbar$.
The number of particles in the pf shell is maintained to be eight ($=7.99$),
but about one more particle is excited from the d$_{5/2}$ 
to the (s$_{1/2}$d$_{3/2}$) subspace. The corresponding configuration
is thus approximated as (d$_{5/2})^{-2}$(d$_{3/2}$s$_{1/2}$)$^{6}$(fp)$^{8}$.

Comparing these two  tables (and also from Fig.\ref{occupation}),
it can be seen that the rotational band is created 
 through two modes. One is excitation within the sd shell,
mainly an excitation from the d$_{5/2}$ orbital
to the upper sd shell (s$_{1/2}$d$_{3/2}$);
the other is within the pf shell and the relevant excitation 
is mainly from the p$_{3/2}$ orbital to the f$_{5/2}$ orbital. 
It is also learned from the calculations that
the numbers of particles in the f$_{7/2}$ and the d$_{3/2}$ orbitals
are almost constant in a wide range of the total angular momentum.

From the above analysis, it can be said that
the main part of the superdeformed structure is determined by the eight
particles in the pf shell. 
Whereas, rotational members of the SD band are mainly produced
by gradual excitations from the d$_{5/2}$ orbital to the upper sd shell
in our model, in addition to the minor internal reconfiguration
inside the pf shell.

\begin{table}[bt]
  \begin{ruledtabular}
    \begin{tabular}{cccccccc}
      Orbital& d$_{5/2}$ &s$_{1/2}$ &d$_{3/2}$ &f$_{7/2}$ &p$_{3/2}$ &
      f$_{5/2}$ &p$_{1/2}$ \\
      \hline
      Proton & 5.11& 1.22& 1.67& 2.68& 0.57& 0.52& 0.23\\
      Neutron& 5.11& 1.22& 1.64& 2.64& 0.61& 0.50& 0.24\\
      \hline
      Total  &10.22& 2.44& 3.31& 5.32& 1.18& 1.02& 0.47\\
    \end{tabular}
  \end{ruledtabular}
  \caption{Occupation numbers of $^{40}$Ca at $J=18\hbar$. 
    The subspace (s$_{1/2}$d$_{3/2}$) is occupied by about six ($=5.75$) 
    particles, while the pf shell is filled with about eight ($=7.99$) 
    particles.     
    The hole occupation number in the d$_{5/2}$ orbital is 1.78 ($=12-10.22$).}
  \label{table2}
\end{table}

\begin{figure}[tb]
  \includegraphics[width=0.35\textwidth,angle=-90]{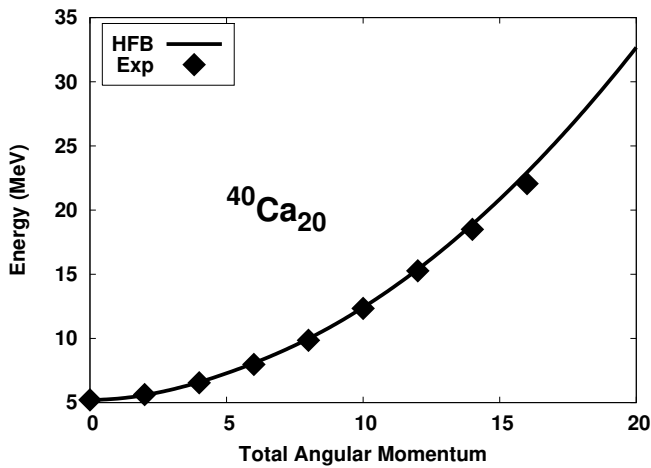}
  \caption{Calculated and observed rotational energy of the SD band 
    in $^{40}$Ca. The calculated ground-state energy (at $J=0$) 
    is normalized with the experimental value, $E(J=0)=5.218$ MeV.}
  \label{SDband}
\end{figure}
\begin{figure}[tbh]
  \includegraphics[width=0.35\textwidth,angle=-90]{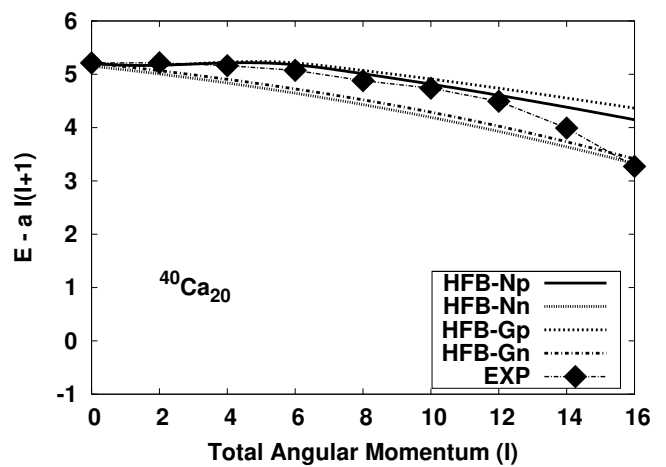}
  \caption{Excitation energies relative to a rigid-rotor energy, 
    $E_{R}=aJ(J+1)$, where $a=0.06909$ (MeV) for $^{40}$Ca.
    The four calculated results (HFB-Np,Nn,Gp, and Gn) 
    are compared with the experimental data.
    The HFB calculations with (without) the g$_{9/2}$ orbital are denoted 
    as G (N). In addition, two different sets of the initial value for the 
    pairing are chosen for each case. The set p corresponds to 
    $(\Delta_{\text{p}}^0,\Delta_{\text{n}}^0)=(1.360,1.513)$ MeV,
    while the set n to $(\Delta_{\text{p}}^0,\Delta_{\text{n}}^0)=(0.15,0.15)$ MeV.
  }
  \label{SDene}
\end{figure}
\begin{figure*}[bth]
  \begin{tabular}{cc}
    \includegraphics[width=0.35\textwidth,angle=-90]{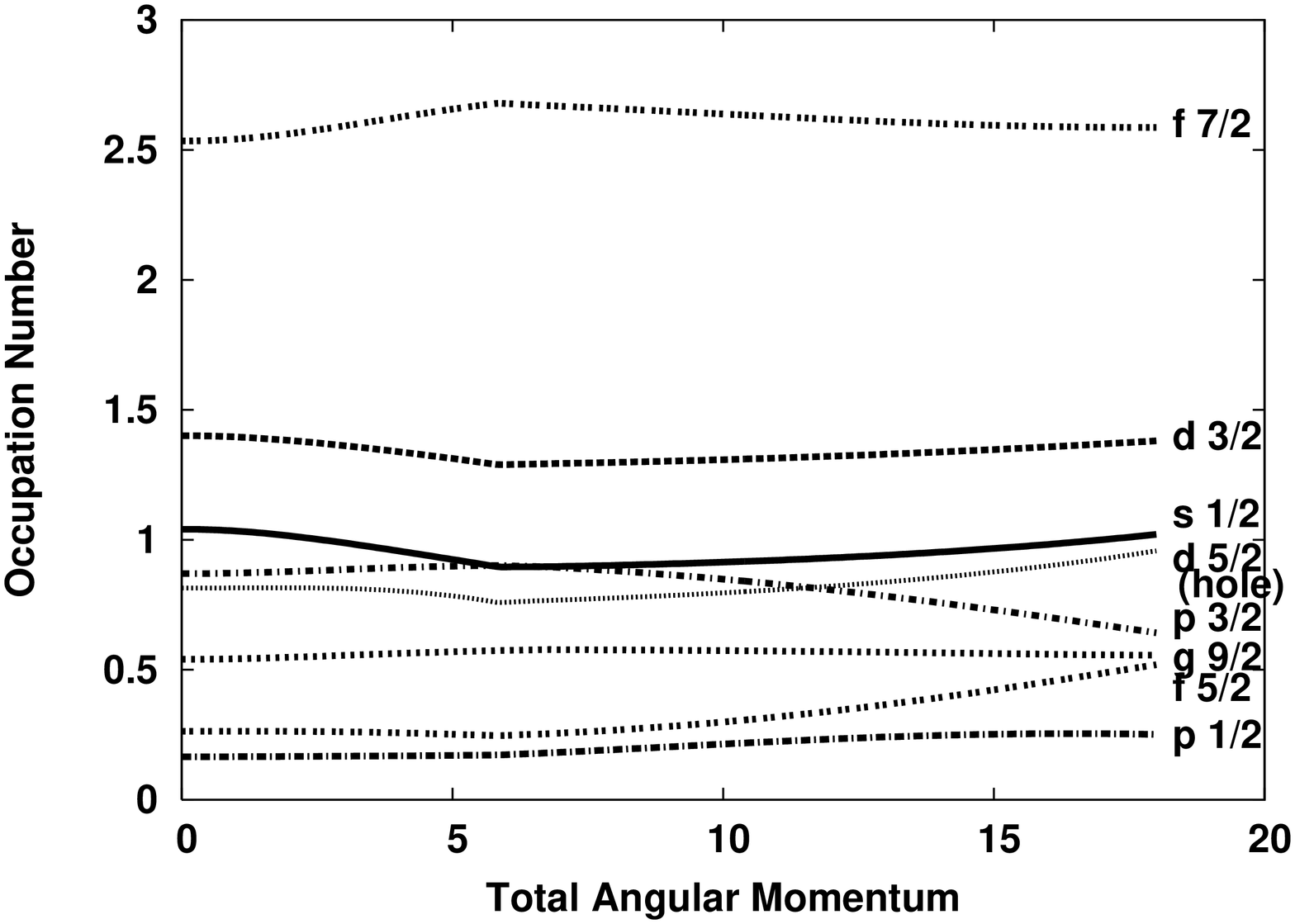}&
    \includegraphics[width=0.35\textwidth,angle=-90]{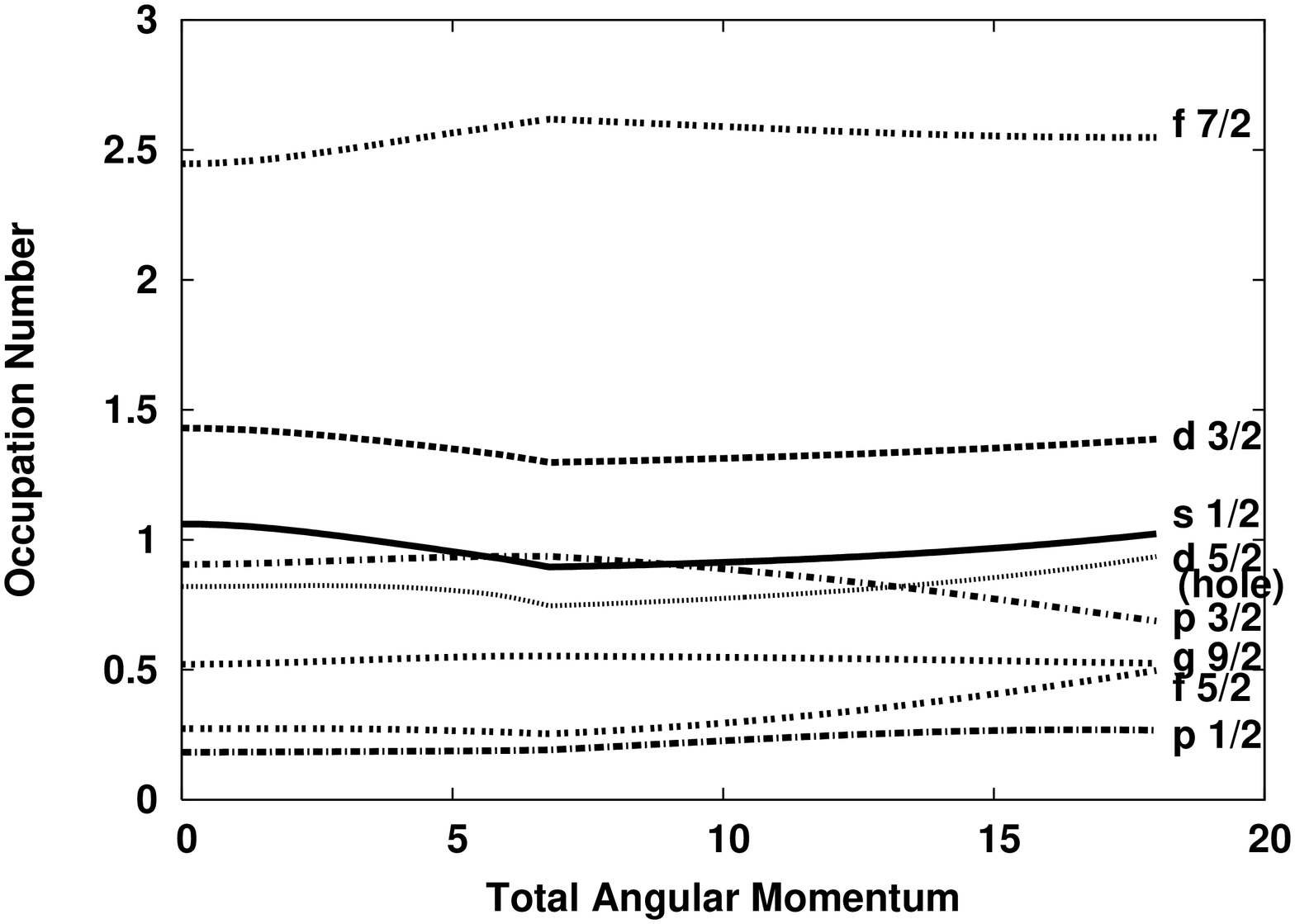}
  \end{tabular}
  \caption{Proton (left) and neutron (right) 
    occupation numbers as a function of the total angular momentum
    for $^{40}$Ca.
    In this case, the g$_{9/2}$ orbital is included in the model
    space.}
  \label{occupation_g}
\end{figure*}

\subsubsection{A role of the g$_{9/2}$ orbital in $J\le 16\hbar$ and rotational energy}

It is worth examining here an effect of the g$_{9/2}$ orbital ($N=4$),
which is missing from the shell model calculation.

Let us see first the calculated rotational energy of the SD band
with the original single-particle model space, 
that is, without the g$_{9/2}$ orbital.
The calculated and observed spectra of the SD band 
are plotted in Fig.\ref{SDband}.
The band-head ($J=0$) energy of the calculated spectrum is normalized
with the experimental data, $E(J=0)=5.218$ MeV. 
The agreement looks very good,
particularly at low spin ($J\alt 12\hbar$).
In Fig.\ref{SDene}, the excitation energy is plotted, 
following Ref.\cite{ISR01}, relative to a rigid rotor energy, 
$E_R=0.06909 \ J(J+1)$ [MeV]
(with a line labeled as HFB-Np, which is
performed in the absence of the g$_{9/2}$ orbital in the model space). 

Despite this good agreement, 
the role of the g$_{9/2}$ orbital is still worth an examination
because the (deformed) Nilsson model implies that some of the split g$_{9/2}$ 
states intrude into the sd shell ($N=2$) at $\beta\simeq 0.6$. 
(See, for example, Fig 2.21a, p.73 in Ref.\cite{RS80}.)
For this purpose, the g$_{9/2}$ orbital 
is added to the model space in the present
framework, and the calculation is repeated with the pairing force 
being unchanged.
The result is plotted with a line denoted as HFB-Gp in Fig. \ref{SDene}.
The low-spin behavior is almost identical to the previous case (HFB-Np),
that is, without the g$_{9/2}$ orbital.
A small deviation from the HFB-Np can be seen at high spin 
($J\agt 12\hbar$), but in practice this difference is negligible
as far as the rotational energy is concerned.
The occupation numbers are also plotted in Fig.\ref{occupation_g}, which 
shows only slight differences from Fig.\ref{occupation}.
It looks that the inclusion of the g$_{9/2}$ orbital gives rise to only a 
minor influence to the nuclear structure, 
but it turns out to be quite essential to
the high-spin nuclear structure of $^{40}$Ca, through the subsequent analyses.
We will come back to this argument in connection to backbending.

\subsubsection{An effect of pairing correlation and backbending}
An effect of the pairing correlation can be also studied here.
Our approach here is to compare two cases: with and without the pairing.
The case without the pairing is constructed by choosing
the initial pairing-gap energies to be small:
$(\Delta^0_{\text{p}},\Delta^0_{\text{n}})=(0.150,0.150)$ [MeV].
With this choice, the gap energies disappear as early as $J\simeq 0.1\hbar$
and remain to do so at higher spin. This calculation is essentially the 
Hartree-Fock (HF) calculation without the pairing, like
the cranked RMF \cite{ISR01} and the cranked Skyrme HF \cite{IMY02}.

\begin{figure}[tbh]
  \includegraphics[width=0.35\textwidth,angle=-90]{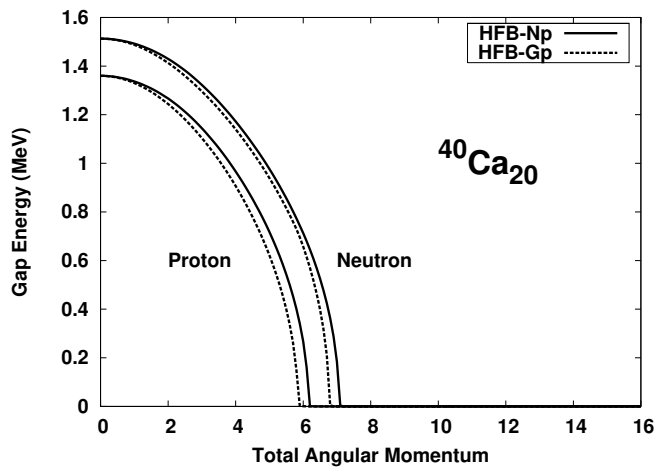}
  \caption{Calculated pairing-gap energies as a function of the total 
  angular momentum for $^{40}$Ca. The identification of the curves
  are the same as Fig.\ref{SDene}.}
\label{gap}
\end{figure}

The rotational energies in Fig.\ref{SDene} denoted HFB-Nn and HFB-Gn 
correspond to the case with and without the g$_{9/2}$ orbital, respectively
(the both cases are without the pairing).
As seen in the figure, there is no much difference between these two cases,
but the both of them underestimate the experimental data. 
When the pairing is adequately taken into account (HFB-Np and HFB-Gp),
there is a plateau structure in the graph at low-spin region, 
which brings a better agreement to the experimental data. 
However, when the pairing correlations are absent,
the plateau structure disappears and the steeper curves appear.
The similar results were obtained in the other mean-field calculations 
neglecting the pairing correlation, 
such as the cranked RMF model \cite{ISR01} 
and the cranked Skyrme HF calculation \cite{IMY02}. 
In the case of $^{36}$Ar,
the shell model calculation \cite{SMJ00} and the PSM \cite{LS01}
reproduce the experimental data fairly well, and the plateau structure
is seen in these calculations. These results suggest the importance of the
higher order correlations in the two-body interaction beyond the mean-field 
level. However, our calculation also implies that an inclusion of
the pp-channels in the mean-field approximation,
that is, the pairing correlations, seem to ``salvage'' effectively
the important correlations that the ph-channels in the mean-field approximation
fail to pick up.
To support this remark, 
discrepancies in the excitation energy start to happen
(which is of the order of about 1 MeV),
after the gap energies disappear at $J\agt 6-7\hbar$ 
(See Figs.\ref{SDene} and \ref{gap}, as well as
the subsequent discussion in the following paragraph).

Fig.\ref{gap} displays the calculated pairing-gap energies for protons and
neutrons with the original initial values for $(\Delta_p^0,\Delta_n^0)
=(1.360,1.513)$ MeV. In either case, the gap energies disappear 
at $J\agt 8\hbar$, which is roughly consistent with the point where
the deviation in the calculated excitation energy can be seen 
in Fig.\ref{SDene}. In finite systems, the pairing correlation should persist
even at high spin, like in the Fig.2(d) in Ref.\cite{LS01}.
This ``collapse'' of the pairing is a notorious problem 
in the BCS-type theory applied to a finite system, 
and this ``phase transition'' of the pairing gap 
is known to be a mere artifact of the model. 
In reality, nuclear systems should undergo a crossover, 
or a gradual decrease in the pairing gap. 
In the present work,
the disappearance of the calculated gap energies is rather smooth and 
gradual, so that the qualitative feature of the system 
might be expected to be maintained, as Ring and Schuck explain 
at p.278 in their textbook \cite{RS80}.
However, a more elaborate treatment to keep the pairing correlation
is necessary for more accurate descriptions at high spin, 
such as the Lipkin-Nogami method \cite{PNL72}.

So far, no backbending is reported in the SD band of $^{40}$Ca
(until $J=16\hbar$),
and our calculation is consistent with this observation
 (with or without an inclusion of the g$_{9/2}$ orbital). 
The shell model calculation by Poves also reproduced this result.
Interestingly, the shell model calculation
 predicts the backbending at higher spin ($J\simeq 20\hbar$) \cite{Po04}.  
 This angular momentum  corresponds to the band termination for
the 8p-8h configuration, that is, (d$_{3/2}$s$_{1/2}$)$^{4}$(f$_{7/2}$)$^{8}$.
According to Poves, the nuclear structure after the backbending
is constructed by such configurations as 
(d$_{3/2}$s$_{1/2}$)$^{4}$(f$_{7/2}$)$^{7}$p$_{3/2}$, as well as 
similar configurations allowing excitations into higher orbitals 
in the pf shells.

In order to see clearly how backbending occurs, 
the so-called ``backbending plot'' is convenient.
In this paper,
the transition energy, $E_{\gamma}(J)$, is defined as
\begin{equation}
  E_{\gamma}(J) \equiv E(J)-E(J-2),
\end{equation}
where $E_{\gamma}(0)=0$. Alternatively, the rotational frequency is
defined as 
\begin{equation}
  \omega(J)=E_{\gamma}(J)/2\hbar,
\end{equation}
and $\omega(0)=0$.
\begin{figure}[tbh]
  \includegraphics[width=0.35\textwidth,angle=-90]{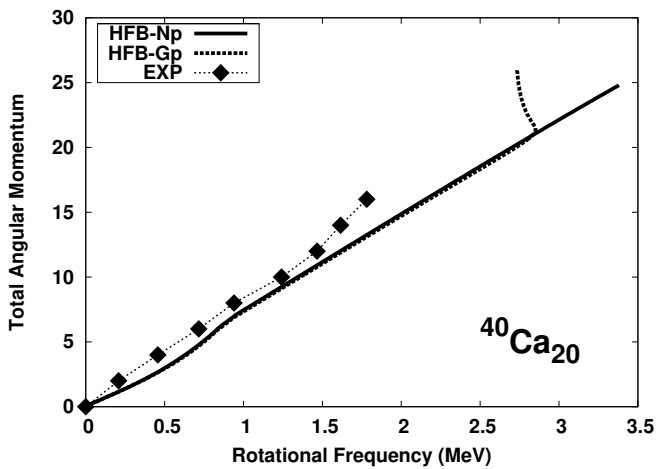}
  \caption{Backbending plot of the SD band of $^{40}$Ca.
  HFB-Np (HFB-Gp) corresponds to the calculation without (with) the
  g$_{9/2}$ orbital in the model space. }
  \label{BB}
\end{figure}

In Fig.\ref{BB}, 
two cases of the calculations are plotted 
(with and without the g$_{9/2}$ orbital). 
The line denoted by HFB-Np (HFB-Gp) corresponds to the case
without (with) the g$_{9/2}$ orbital in the model space.
As confirmed in Figs.\ref{occupation} and \ref{occupation_g},
there is no much difference between these two cases in the spin range
$J\alt 20\hbar$. This situation is reflected in Fig.\ref{BB} showing
completely the same behaviors in the two calculated lines 
in $J\alt 20\hbar$.
A difference happens beyond $J=20\hbar$.
The case without the g$_{9/2}$ orbital (HFB-Np) shows no sign of backbending
even at as high as $J=25\hbar$ \footnote{The calculation of HFB-Np can be
executed up to $J\simeq 25\hbar$. Beyond this angular momentum, the SD structure
no longer exists in the current framework. 
The cranked Skyrme HF calculation by Inakura, et al. also shows
that the SD structure ends at $J=24\hbar$.},
while the line of HFB-Gp starts to backbend at $J=20\hbar$, as
predicted by the shell model calculation. However, it should be noted that
the shell model calculation does not contain 
the g$_{9/2}$ and d$_{5/2}$ orbitals.

\subsubsection{A role of the g$_{9/2}$ orbital in the backbending}

To study the rotational alignment, it is useful to calculate
the single-particle angular momentum component along the cranking axis.
The quantity is given as
\begin{equation}
  \langle{j}_x(m)\rangle = \sum_{n}\rho_{mn}(j_x)_{nm},
\end{equation}
where the indices $m,n$ denote the spherical Nilsson basis.

\begin{figure*}[htb]
  \begin{tabular}{cc}
    \includegraphics[width=0.35\textwidth,angle=-90]{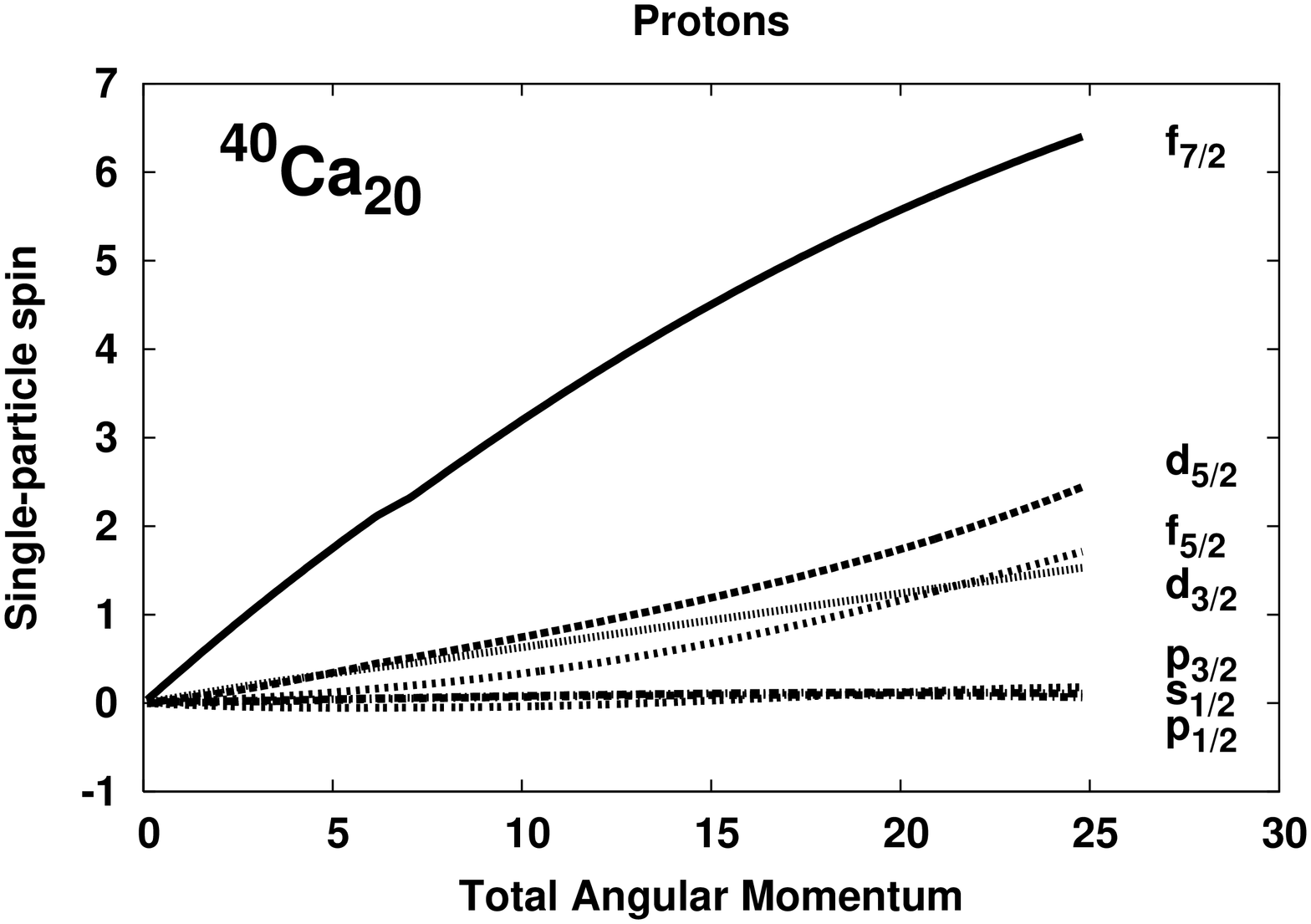}&
    \includegraphics[width=0.35\textwidth,angle=-90]{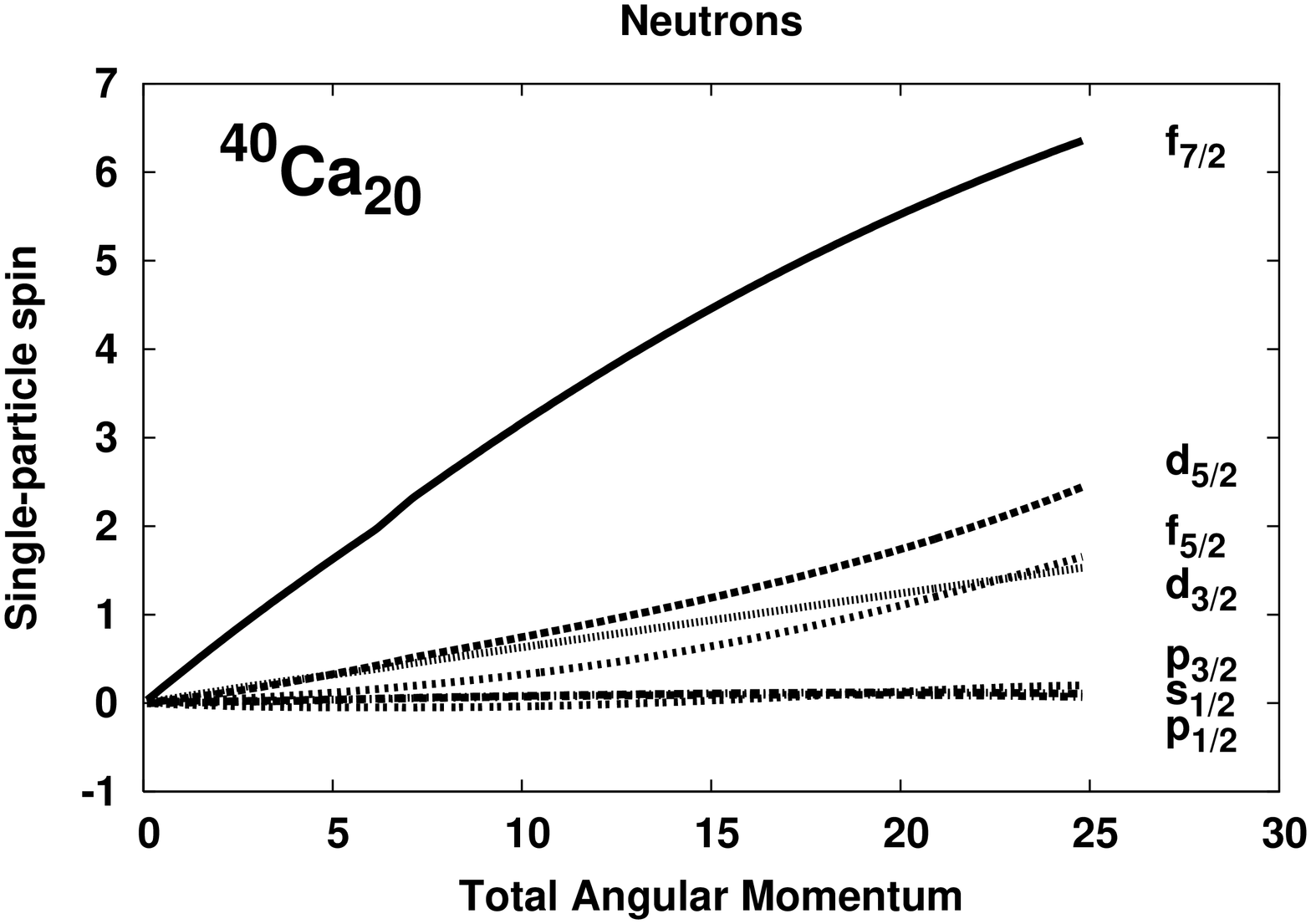}
  \end{tabular}
  \caption{Proton (left) and neutron (right) 
    single-particle spin components along the cranking axis
    for $^{40}$Ca. 
    In this case, the g$_{9/2}$ orbital is excluded from the model
    space.}
  \label{algn_n}
\end{figure*}

\begin{figure*}[htb]
  \begin{tabular}{cc}
    \includegraphics[width=0.35\textwidth,angle=-90]{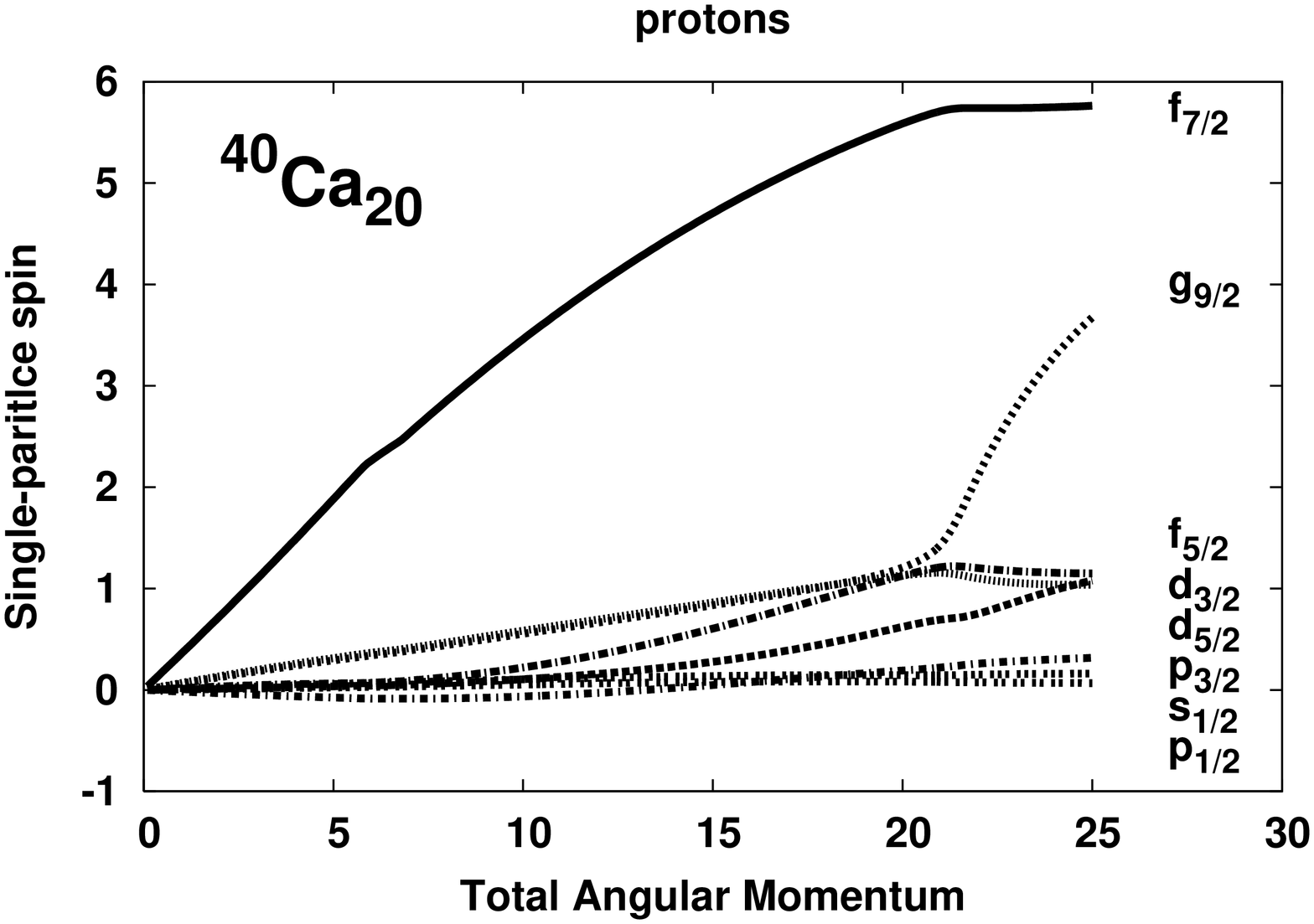}&
    \includegraphics[width=0.35\textwidth,angle=-90]{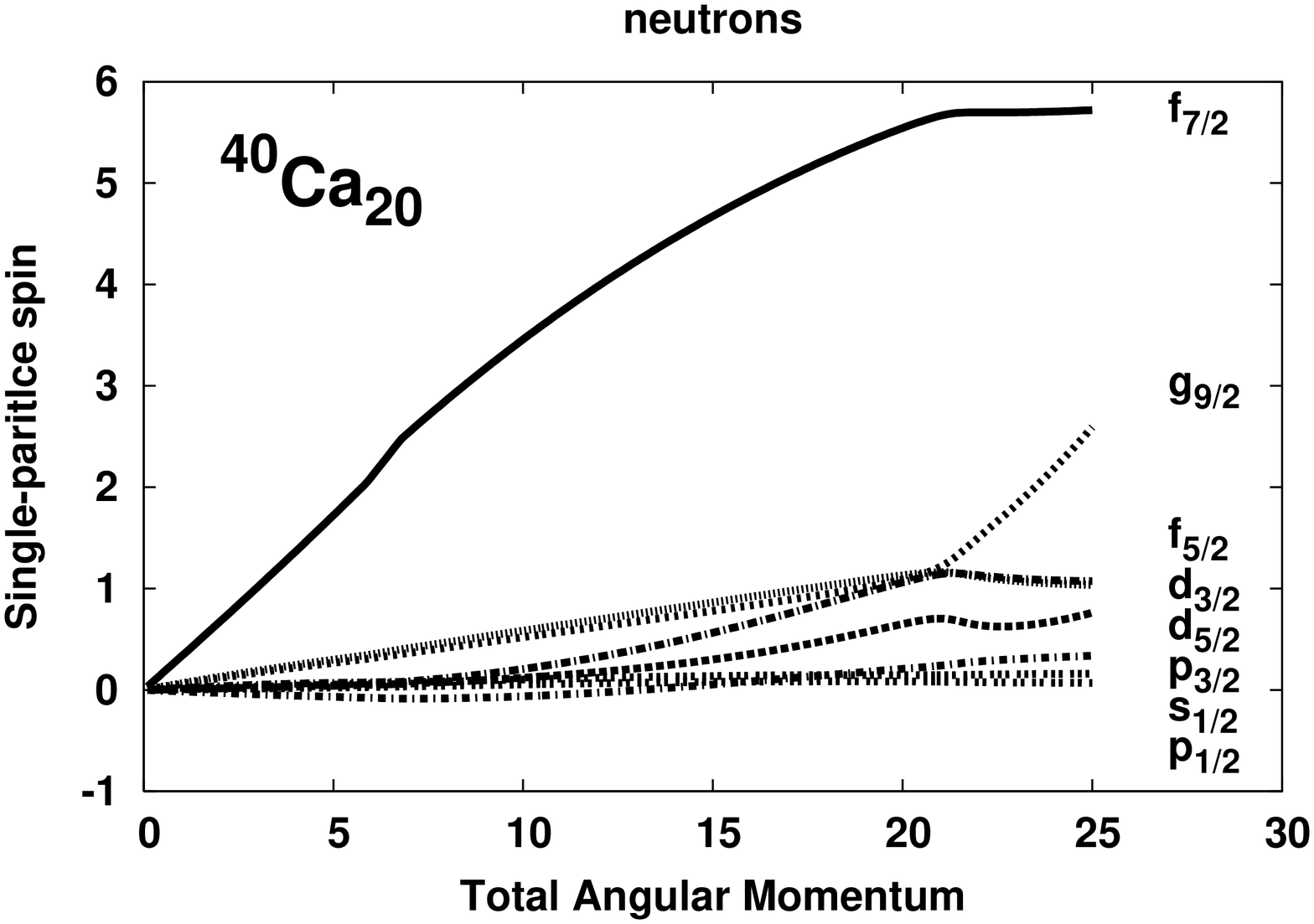}
  \end{tabular}
  \caption{Proton (left) and neutron (right) 
    single-particle spin components along the cranking axis
    for $^{40}$Ca. 
    In this case, the g$_{9/2}$ orbital is included in the model
    space.}
  \label{algn_g}
\end{figure*}

Fig.\ref{algn_n} shows the single-particle spin component along the
cranking axis, in the case where the g$_{9/2}$ orbital 
is excluded from the model space.
The f$_{7/2}$ orbital is the major contributor to the total angular momentum, 
and its contribution becomes gradually increased at higher spin states.
About 60\% of the total angular momentum is produced by this orbital
(of protons and neutrons).
The second main component is produced by the d$_{5/2}$ orbitals, and its
percentage to the total angular momentum reaches nearly 20\% at $J=25\hbar$. 
The contributions from the d$_{3/2}$ and f$_{5/2}$ orbitals 
are also significant at high spin.

Fig.\ref{algn_g} shows the single-particle spin components 
when the g$_{9/2}$ orbital is included in the model space.
Beyond $J=20\hbar$, a clear structural change is seen. That is, 
the contributions from the f$_{7/2}$ orbital as well as d$_{3/2}$ and f$_{5/2}$ 
orbitals are saturated. Instead, a rapid alignment of the g$_{9/2}$ orbital 
happens. Obviously, the backbending seen in Fig.\ref{BB} 
is caused by this structural change.
As mentioned earlier, in the case of the shell model calculation, 
backbending is initiated by the band termination, 
(d$_{3/2}$s$_{1/2}$)$^{4}$(f$_{7/2}$)$^{8}$. The higher spin states are
produced as excitations happen from the f$_{7/2}$ orbital to higher orbitals 
in the pf shell. Comparing with our results,
the backbending occurs from similar reasons, but not exactly the same.
First of all, the direct factor to cause the backbending is the rapid
alignment in the $g_{9/2}$ orbital, 
which corresponds to an excitation to higher orbitals, 
but this excitation is more drastic in the sense that three major shells
are involved (i.e., sd-pf-g). In addition, the saturation of the alignment 
in the f$_{7/2}$ orbital produces the similar mechanism as the band termination.
In fact, the maximum angular momentum generated by two particles occupying 
the f$_{7/2}$ orbital is $6\hbar$, 
which is almost the value read in Fig.\ref{algn_g}
for the f$_{7/2}$ orbital. 
Unlike the simple band termination picture, 
the d$_{3/2}$ and f$_{5/2}$ orbitals also show the saturation,
despite that the generated angular momenta are less than the maximum values.
No more additional angular momentum is created by further alignments 
of these orbitals in the sd and pf shells beyond $J=20\hbar$ 
(i.e., before the backbending). 
In other words, beyond this total angular momentum,
only one high-$j$ orbital (g$_{9/2}$) dominantly produces an additional
angular momentum on top of the collective angular momentum already produced by
the particles in the sd and pf shells. This mechanism is exactly the same
as the original backbending mechanism in the rare-earth nuclei, 
where the i$_{13/2}$ orbital (usually, of neutrons) plays the same role 
as its counterpart, that is, the g$_{9/2}$ orbital.
\begin{table}[bt]
  \begin{ruledtabular}
    \begin{tabular}{ccccccccc}
      Orbital& d$_{5/2}$ &s$_{1/2}$ &d$_{3/2}$ &f$_{7/2}$ &p$_{3/2}$ &
      f$_{5/2}$ &p$_{1/2}$ & g$_{9/2}$\\
      \hline
      Proton & 5.04& 1.02& 1.38& 2.59& 0.64& 0.52& 0.25& 0.56\\
      Neutron& 5.06& 1.02& 1.39& 2.55& 0.69& 0.50& 0.27& 0.53\\
      \hline
      Total  &10.10& 2.04& 2.77& 5.14& 1.33& 1.02& 0.52& 1.09\\
    \end{tabular}
  \end{ruledtabular}
  \caption{Occupation numbers of $^{40}$Ca at $J=18\hbar$ in the case
    with the g$_{9/2}$ orbital included in the model space. 
    The subspace (s$_{1/2}$d$_{3/2}$) is occupied by about five ($=4.81$) 
    particles, while the pf shell is filled with about eight ($=8.01$) 
    particles.     
    The hole occupation number in the d$_{5/2}$ is 1.90 ($=12-10.10$).}
  \label{table3}
\end{table}

\begin{table}[bt]
  \begin{ruledtabular}
    \begin{tabular}{ccccccccc}
      Orbital& d$_{5/2}$ &s$_{1/2}$ &d$_{3/2}$ &f$_{7/2}$ &p$_{3/2}$ &
      f$_{5/2}$ &p$_{1/2}$ & g$_{9/2}$\\
      \hline
      Proton & 4.46& 1.00& 1.39& 2.49& 0.66& 0.59& 0.26& 1.16\\
      Neutron& 4.65& 0.99& 1.40& 2.45& 0.71& 0.56& 0.28& 0.96\\
      \hline
      Total  & 9.11& 1.99& 2.79& 4.94& 1.37& 1.15& 0.54& 2.12\\
    \end{tabular}
  \end{ruledtabular}
  \caption{Occupation numbers of $^{40}$Ca at $J=26\hbar$ in the case
    with the g$_{9/2}$ orbital included in the model space. 
    The subspace (s$_{1/2}$d$_{3/2}$) is occupied by about five ($=4.78$) 
    particles, while the pf shell is filled with eight ($=8.00$) 
    particles.     
    The hole occupation number in the d$_{5/2}$ is 2.89 ($=12-9.11$).}
  \label{table4}
\end{table}
The calculated occupation numbers (Tables \ref{table3} and \ref{table4}) 
imply that the configuration changes
from (d$_{5/2}$)$^{-2}$(d$_{3/2}$s$_{1/2}$)$^{5}$(fp)$^{8}$(g$_{9/2}$)$^{1}$
to (d$_{5/2}$)$^{-3}$(d$_{3/2}$s$_{1/2}$)$^{5}$(fp)$^{8}$(g$_{9/2}$)$^{2}$ in the backbending region.
Essentially, this change is brought by an excitation from the d$_{5/2}$ orbital
to the g$_{9/2}$ orbital, 
while the configurations inside the pf shell and the subspace of the
sd shell are relatively stable throughout the whole  range of angular momentum.
This excitation from the d$_{5/2}$ orbital to the g$_{9/2}$ orbital 
can be understood 
through the Nilsson diagram around $\beta\simeq 0.6$. 
There, a low-$\Omega$ component originating from the
g$_{9/2}$ orbital, that is, $[440]_{1/2}$, behaves like an intruder orbital
coming down to the region near the d$_{5/2}$ and d$_{3/2}$ orbitals. 
This situation implies that these three positive-parity states
can jointly compose the nuclear many-body state
when the system undergoes superdeformation.
To produce high angular momentum, it is efficient to place more particles
into the g$_{9/2}$ orbital. 

\begin{table}[bt]
  \begin{ruledtabular}
    \begin{tabular}{cccccccc}
      Orbital& d$_{5/2}$ &s$_{1/2}$ &d$_{3/2}$ &f$_{7/2}$ &p$_{3/2}$ &
      f$_{5/2}$ &p$_{1/2}$ \\
      \hline
      Proton & 5.02& 1.29& 1.70& 2.70& 0.50& 0.59& 0.21\\
      Neutron& 5.02& 1.29& 1.70& 2.66& 0.54& 0.56& 0.23\\
      \hline
      Total  &10.04& 2.58& 3.40& 5.36& 1.04& 1.15& 0.44\\
    \end{tabular}
  \end{ruledtabular}
  \caption{Occupation numbers of $^{40}$Ca at $J=20\hbar$ in the case
    with the g$_{9/2}$ orbital excluded from the model space. 
    The subspace (s$_{1/2}$d$_{3/2}$) is occupied by about six ($=5.98$) 
    particles, while the pf shell is filled with about eight ($=7.99$) 
    particles.     
    The hole occupation number in the d$_{5/2}$ orbital is 1.96 ($=12-10.04$).}
  \label{table5}
\end{table}
\begin{table}[bt]
  \begin{ruledtabular}
    \begin{tabular}{cccccccc}
      Orbital& d$_{5/2}$ &s$_{1/2}$ &d$_{3/2}$ &f$_{7/2}$ &p$_{3/2}$ &
      f$_{5/2}$ &p$_{1/2}$\\
      \hline
      Proton & 4.78& 1.46& 1.77& 2.75& 0.35& 0.73& 0.17\\
      Neutron& 4.78& 1.46& 1.77& 2.72& 0.39& 0.71& 0.18\\
      \hline
      Total  & 9.56& 2.92& 3.54& 5.47& 0.74& 1.44& 0.35\\
    \end{tabular}
  \end{ruledtabular}
  \caption{Occupation numbers of $^{40}$Ca at $J=26\hbar$ in the case
    with the g$_{9/2}$ orbital excluded from the model space. 
    The subspace (s$_{1/2}$d$_{3/2}$) is occupied by about five ($=6.46$) 
    particles, while the pf shell is filled with eight ($=8.00$) 
    particles.     
    The hole occupation number in the d$_{5/2}$ orbital is 2.44 ($=12-9.56$).}
  \label{table6}
\end{table}

\begin{figure*}[tbh]
  \begin{tabular}{cc}
    \includegraphics[width=0.35\textwidth,angle=-90]{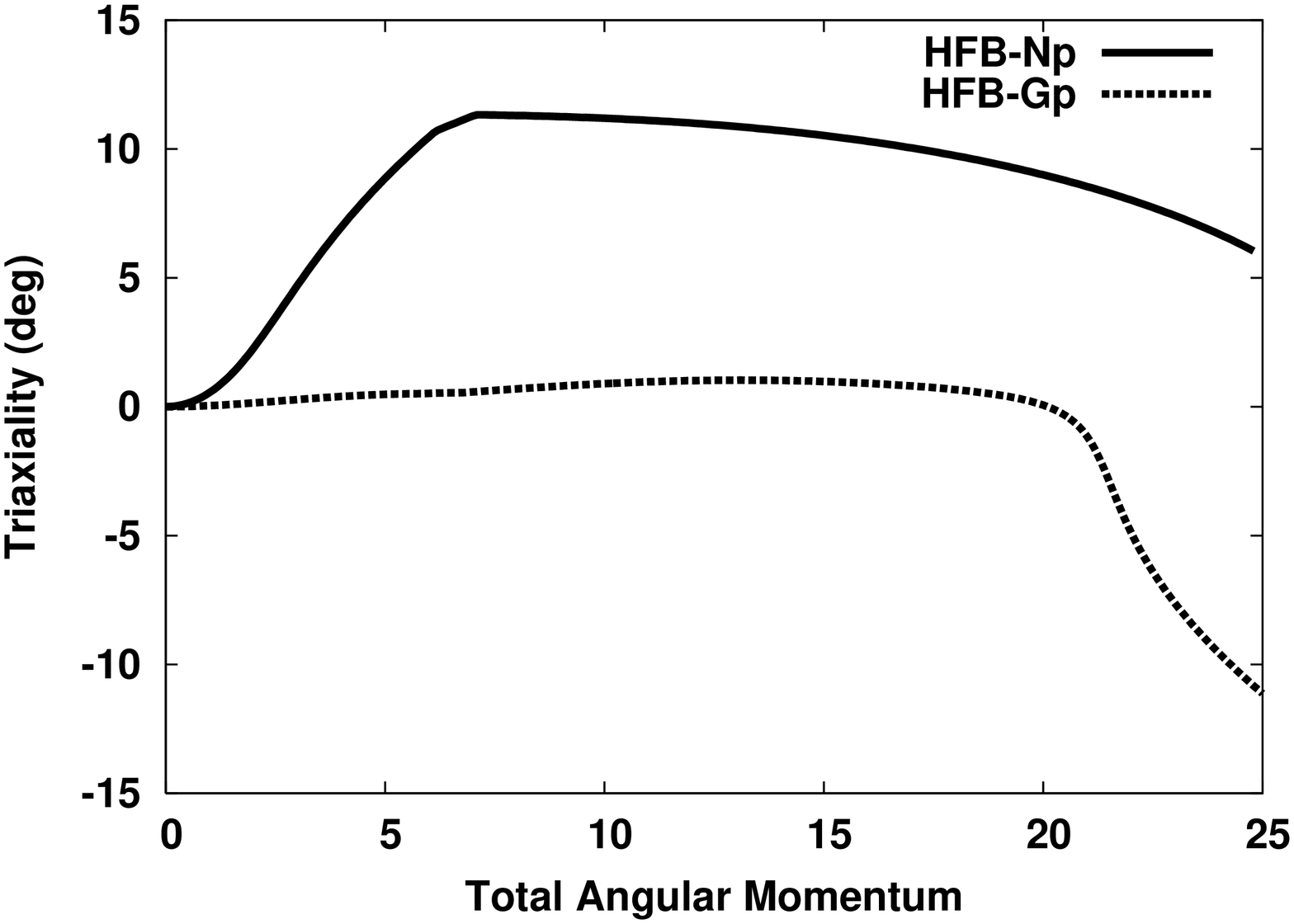}&
    \includegraphics[width=0.35\textwidth,angle=-90]{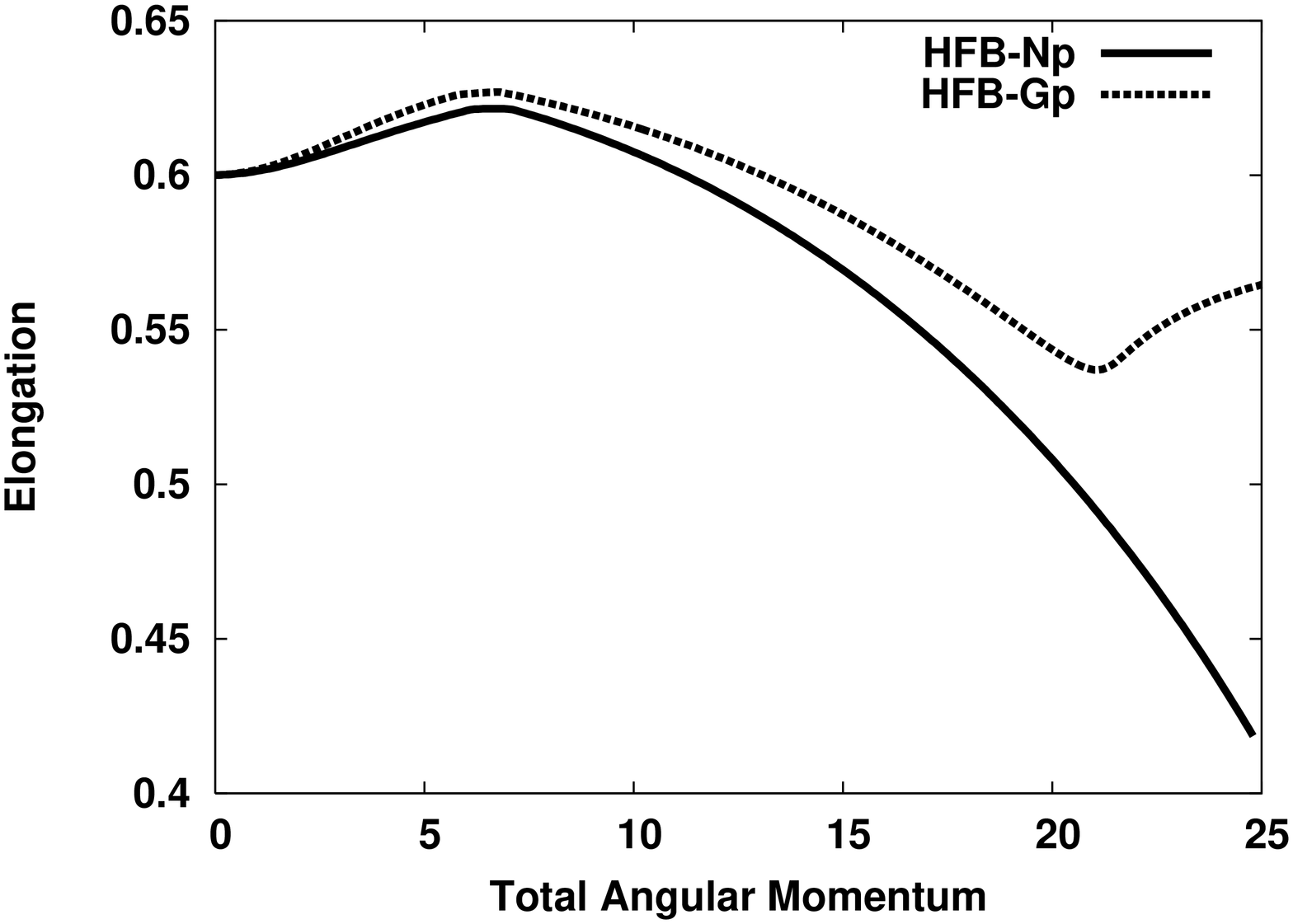}
  \end{tabular}
  \caption{Calculated triaxiality $\gamma$ (left panel) 
    and elongation $\beta$ (right), as functions of the total
    angular momentum  for $^{40}$Ca. Captions {HFB-GP} and { HFB-NP}
    denote the cases with and without the g$_{9/2}$ orbital, respectively.}
  \label{Qdef}
\end{figure*}

It is worth studying more closely the case without the g$_{9/2}$ orbital, 
in which our calculation shows no sign of backbending.
From Tables \ref{table5} and \ref{table6},
the configuration at high spin looks quite stable to keep the
structure of 
(d$_{5/2}$)$^{-2}$(d$_{3/2}$s$_{1/2}$)$^{6}$(fp)$^{8}$.
The rotational members in the band are produced through gradual
excitations from the d$_{5/2}$ orbital to the upper sd shell as well
as a minor rearrangement within the pf shell. In other words,
this structural change is adiabatic against the increment of the Coriolis
force. This adiabatic nature of the high-spin nuclear structure 
is nothing but to mean the regularity of the band. In this way,
no backbending is seen in this case. 
Fig.\ref{algn_n} shows
that the total angular momentum is created mainly through 
the gradual and monotonic alignment in the f$_{7/2}$ orbital. 
In addition, the contributions of the d$_{3/2}$ and the f$_{5/2}$ orbitals
are seen to be non-negligible, as shown in the shell model calculation.
As in the previous case, the d$_{5/2}$ orbital
also contributes to the creation of the total angular momentum.

\subsubsection{Evolution of shape}

Finally, let us examine how shape evolution happens in our calculation.
To show quadrupole deformation, the Hill-Wheeler coordinates ($\beta,\gamma$)
\cite{HW52} are used in this work. That is, 
$\langle\hat{Q}_0\rangle \propto \beta\cos\gamma$ and
$\langle\hat{Q}_2\rangle \propto \beta\sin\gamma/\sqrt{2}$.
In the P$+$Q$\cdot$Q model, the proportional constant carrying the
unit of the quadrupole moment is
given as $\left(\hbar\omega/\hbar c\right)^2 mc^2 \chi^{-1}$, 
where the harmonic oscillator energy is $\hbar\omega = 41A^{-1/3}$ (MeV),
and $\hbar c$=197 (MeV$\cdot$fm).
The mass for a nucleon is $mc^2\simeq 1$ (GeV) and 
the coupling constant $\chi$ is given in Eq.(\ref{QQc}).

Fig.\ref{Qdef} presents  the calculated triaxiality ($\gamma$) and
elongation ($\beta$) for the two cases (with and without
the g$_{9/2}$ orbital, 
which are respectively denoted as HFB-Gp and HFB-Np, in the figure). 
Until $J=20\hbar$, the elongation does not change significantly.
The higher the total angular momentum, the more shrinkage of the deformation
can be seen along the longest principal axis of the quadrupole moment.
But the two curves start to deviate from each other 
beyond the angular momentum $J=20\hbar$.
The curve corresponding to the case without the g$_{9/2}$ orbital (HFB-Np)
shows a monotonic decrease to reach $\beta\simeq 0.42$, 
while the curve corresponding to
the case with the g$_{9/2}$ orbital (HFB-Gp) stops decreasing at $J=20\hbar$
to maintain the deformation larger than $\beta=0.55$.
The shell model calculation by Poves \cite{Po04} also implies the shrink
of the shape until $J=18\hbar$ 
in the calculation of the intrinsic quadrupole moment, $Q_0$.
The cranked Skyrme-HF calculation by Inakura et al. \cite{IMY02}
also shows the shrink, and 
the elongation is demonstrated to keep $\beta\agt 0.5$ (for $J\le 24\hbar$)
for the three parameter sets (SIII, SkM$^*$, SLy4).
This result is consistent with our HFB-Gp case, 
that is, the case with the g$_{9/2}$ orbital (and with the non-vanishing
initial pairing-gap parameters).

On the contrary to the elongation, 
there is a significant difference in the triaxiality between the
two cases (i.e., with and without the g$_{9/2}$ orbital).
The case without the g$_{9/2}$ orbital 
shows that the triaxiality is always positive
($\gamma>0^{\circ}$) and the triaxial deformation already starts to grow 
at low spin although the triaxiality is not so large 
($0^{\circ}\alt\gamma\alt 10^{\circ}$).
The other case (with the g$_{9/2}$ orbital) 
shows that the nucleus is axially symmetric
until the backbending starts at $J=20\hbar$. Beyond the backbending 
angular momentum ($J=20\hbar$), the triaxiality quickly starts to grow 
with negative values,
which are consistent with the picture of the band termination.
The amount of triaxiality is, however, not so substantial 
($|\gamma|\alt 10^{\circ})$ even at $J\simeq 25\hbar$.
The cranked Skyrme-HF calculation \cite{IMY02} gave the consistent results with
our calculation for the case with the g$_{9/2}$ orbital,
 that is, the $\gamma$ is negative and the amount 
(an absolute value of $\gamma$) is less than $10^{\circ}$ for $J\le 24\hbar$.

\subsubsection{Summary for $^{40}$Ca}
Let us summarize here our analysis on the SD states of $^{40}$Ca.
The structure of the SD band of $^{40}$Ca is mainly determined
by the eight particles placed in the pf shell. 
About five to six particles sit in the (s$_{1/2}$d$_{3/2}$) subspace 
of the sd shell, which corresponds to the configuration of six to seven holes. 
The truncation of the d$_{5/2}$ orbital
in the shell model calculation demands the eight-hole configuration in the
subspace, which is approximately consistent with our result.
However, for more accurate descriptions, 
the d$_{5/2}$ and g$_{9/2}$ orbitals need to be taken into account
in the model space. This is particularly so for the description 
of the nuclear structure at high spin ($J\agt 20\hbar$),
where the backbending is predicted. 

In the presence of these additional orbitals,
the phenomenon similar to the band termination starts 
to occur at $J=20\hbar$ in our model,
but the saturation of alignments in the pf and sd shells are assisted
by the quick alignment in the g$_{9/2}$ orbital, to which particles are excited
from the d$_{5/2}$ orbital.  This mechanism is consistent with
 the deformed Nilsson model, where the g$_{9/2}$ orbital comes down to the
sd shell, as an ``intruder'' orbital at superdeformation ($\beta\simeq 0.6$). 
As a result of the saturation of the spin alignment in the sd-pf shell
and the quick alignment of the g$_{9/2}$ orbital, 
triaxial deformation starts occur,
but its amount is not so substantial ($|\gamma|\alt 10^{\circ}$)
that one can conclude that the nucleus keeps near-axial symmetry
with superdeformation ($\beta\simeq 0.6$).

\subsection{A case of backbending : $^{36}$Ar}

The second targeted nucleus in this paper, $^{36}$Ar, shows 
a backbending at $J=10\hbar$ in its superdeformed band, 
which is currently identified up to $J=16\hbar$ \cite{SMJ00}. 
The (s$_{1/2}$d$_{3/2}$)$^{4}$(pf)$^{4}$ structure is proposed 
in the shell model calculation truncating the d$_{5/2}$ orbital \cite{SMJ00}.
For the cause of the backbending, simultaneous alignments 
of protons and neutrons in the f$_{7/2}$ orbitals were suggested 
by the PSM \cite{LS01}.
\subsubsection{Deformation of the band head}

There are slight disagreements in the calculated band-head
deformation among different models.
In the calculation with the cranked Nilsson model,  
minimization of the potential energy surface $E(\gamma,\beta)$
gave $\beta\simeq 0.45$ and $\gamma=0^{\circ}$ for the band head \cite{SMJ00}. 
The PSM calculation assumed axial symmetry ($\gamma=0^{\circ}$)
and the fixed value $\beta\simeq 0.48$,
no matter how high (or low) the total angular momentum is \cite{LS01}.
The calculations with the cranked Skyrme HF (the SIII and SkM$^*$ 
parameterizations for the interaction \footnote{The SLy4 parameter results in triaxial deformation for the SD bandhead, but this result may be ``less reliable'' than the other parameter sets according to the authors \cite{IMY02}.}) 
resulted in $\beta\simeq 0.5$ and $\gamma=0^{\circ}$ \cite{IMY02}.
In the calculation done by Bender, et al. assuming axial symmetry\cite{BHR03},
there was no SD minimum found in their mean-field solution 
(The Skyrme HF+BCS with the SLy6 force parameterization), 
but the projected solutions gave rise to the minimum at $\beta\simeq 0.5$.

\begin{figure}[tbh]
  \includegraphics[width=0.36\textwidth,angle=-90]{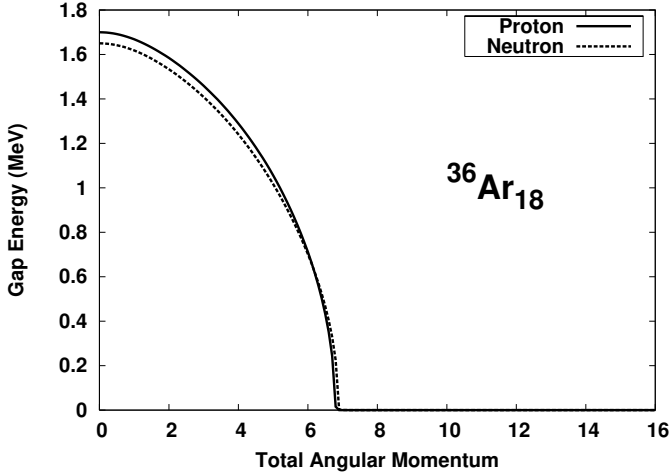}
  \caption{The pairing-gap energies at the total angular momentum 
    $0\le J\le 16\hbar$. The g$_{9/2}$ orbital is included in the model space.}
  \label{gap_Ar36}
\end{figure}

Considering these calculations, the initial deformation parameter 
$\beta^0=0.5$, as well as axial symmetry ($\gamma^0 = 0^{\circ}$), 
seems reasonable for our initial deformation parameters. 
The initial pairing-gap energies are employed to be 
$(\Delta_{\text{p}}^0,\Delta_{\text{n}}^0)=(1.70,1.65)$ MeV,  which
are about 6\% stronger than the values suggested in
Ref.\cite{MN95}. This adjustment is made so as to avoid the sudden 
disappearance of the gap energy. 
With this slight modification, the gap energy turns to
disappear more gradually and smoothly at $J\simeq 7\hbar$,
as shown in Fig.\ref{gap_Ar36}.
We have confirmed, however, that the overall qualitative nature of the SD state
are unchanged for the modification.

\subsubsection{Backbending and roles of the g$_{9/2}$ orbital}

\begin{figure}[tbh]
  \includegraphics[width=0.35\textwidth,angle=-90]{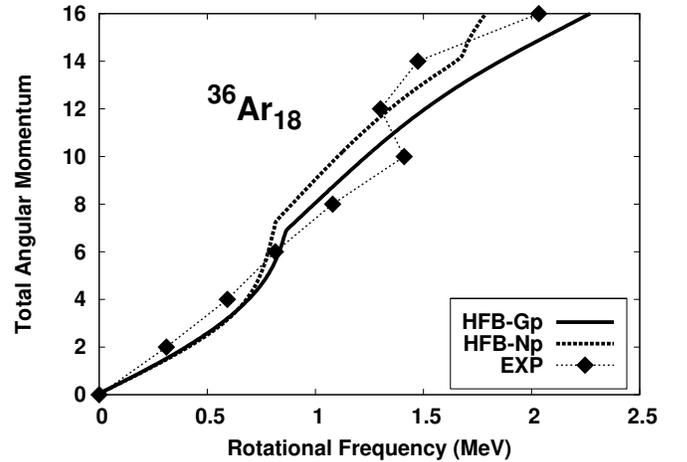}
  \caption{Backbending plot of the SD band of $^{36}$Ar.
  HFB-Np (HFB-Gp) corresponds to the calculation without (with) the
  g$_{9/2}$ orbital in the model space. }
  \label{BB_Ar36}
\end{figure}

To examine the quality of our calculation,
it is useful to see the backbending plot first, which is presented
in Fig.\ref{BB_Ar36}.
Due to the disappearance of the pairing gap at $J\simeq 7\hbar$,
backbending starts earlier (at $J\simeq 7\hbar$) than the experiment
(at $J\simeq 10\hbar$). Having accepted this discrepancy, the calculation
manages to reproduce the qualitative behavior of the backbending profile
of the SD band in this nucleus. 
Hence, in order to discuss the structural change causing the backbending,
it is sufficient to study
the configurations before and after $J\simeq 7\hbar$ in our model.
Let us see the corresponding occupation numbers next.

\begin{figure*}[bth]
  \begin{tabular}{cc}
    \includegraphics[width=0.35\textwidth,angle=-90]{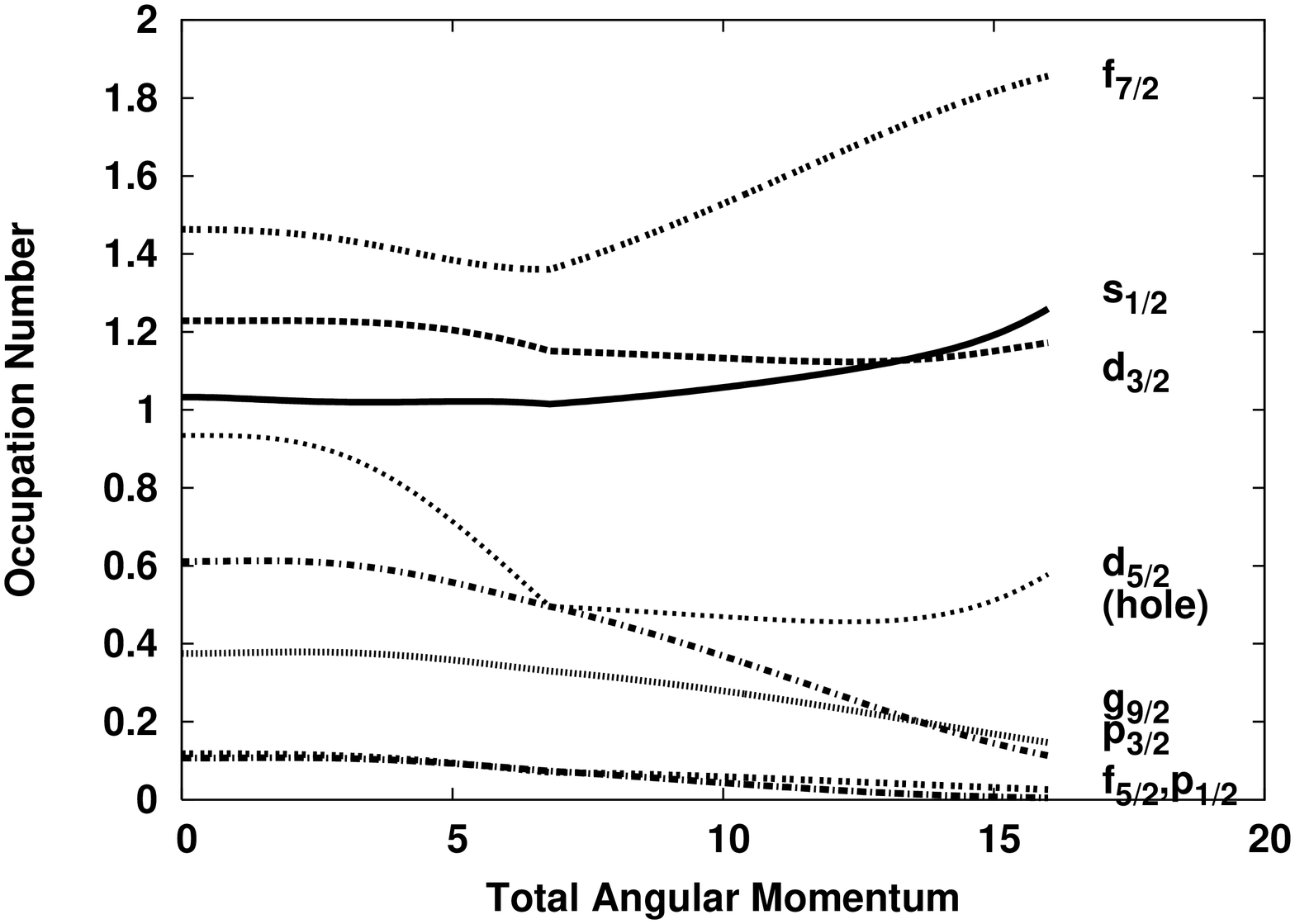}&
    \includegraphics[width=0.35\textwidth,angle=-90]{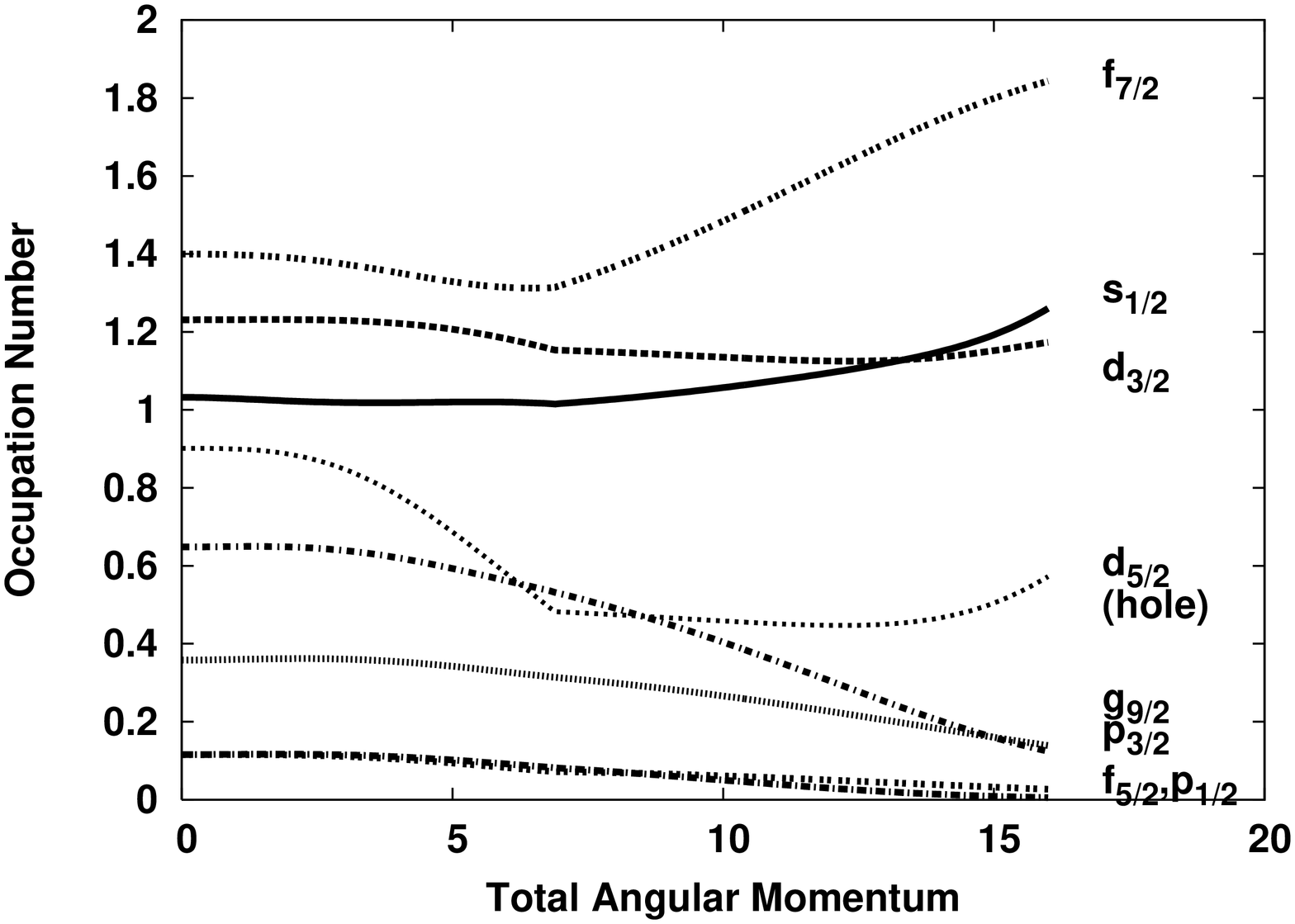}
  \end{tabular}
  \caption{Proton (left) and neutron (right) 
    occupation numbers as a function of the total angular momentum
    for $^{36}$Ar.
    In this case, the g$_{9/2}$ orbital is included in the model
    space.}
  \label{occupation_gAr}
\end{figure*}
\begin{table}[bt]
  \begin{ruledtabular}
    \begin{tabular}{ccccccccc}
      Orbital& d$_{5/2}$ &s$_{1/2}$ &d$_{3/2}$ &f$_{7/2}$ &p$_{3/2}$ &
      f$_{5/2}$ &p$_{1/2}$& g$_{9/2}$\\
      \hline
      Proton & 5.07& 1.03& 1.23& 1.46& 0.61& 0.12& 0.11& 0.38\\
      Neutron& 5.07& 1.03& 1.23& 1.40& 0.65& 0.12& 0.12& 0.36\\
      \hline
      Total  & 10.17& 2.06& 2.46& 2.86& 1.26& 0.24& 0.23& 0.74\\
    \end{tabular}
  \end{ruledtabular}
  \caption{Occupation numbers of $^{36}$Ar at $J=0$ in the case
    with the g$_{9/2}$ orbital included in the model space. 
    The subspace (s$_{1/2}$d$_{3/2}$) is occupied by about four to five 
    ($=4.52$) particles, 
    while the pf shell is filled with four to five ($=4.59$) particles.     
    The hole occupation number in the d$_{5/2}$ orbital is 1.83 ($=12-10.17$).}
  \label{tAr1}
\end{table}
\begin{table}[bt]
  \begin{ruledtabular}
    \begin{tabular}{ccccccccc}
      Orbital& d$_{5/2}$ &s$_{1/2}$ &d$_{3/2}$ &f$_{7/2}$ &p$_{3/2}$ &
      f$_{5/2}$ &p$_{1/2}$& g$_{9/2}$\\
      \hline
      Proton & 5.42& 1.26& 1.17& 1.86& 0.11& 0.03& 0.00& 0.15\\
      Neutron& 5.43& 1.26& 1.17& 1.84& 0.12& 0.03& 0.01& 0.14\\
      \hline
      Total  & 10.85& 2.52& 2.34& 3.70& 0.23& 0.06& 0.01& 0.29\\
    \end{tabular}
  \end{ruledtabular}
  \caption{Occupation numbers of $^{36}$Ar at $J=16\hbar$ in the case
    with the g$_{9/2}$ orbital included in the model space. 
    The subspace (s$_{1/2}$d$_{3/2}$) is occupied by about five ($=4.86$) 
    particles, while the pf shell is filled with four ($=4.00$) 
    particles.     
    The hole occupation number in the d$_{5/2}$ orbital is 1.13 ($=12-10.85$).}
  \label{tAr2}
\end{table}

The occupation numbers are displayed in Fig.\ref{occupation_gAr},
as well as in Tables \ref{tAr1} and \ref{tAr2}.
Table \ref{tAr1}, which presents the configurations before the
backbending, shows that approximately four and a half particles occupy the 
subspace of the sd shell, as well as the pf shell. The g$_{9/2}$ orbital
is occupied by less than one particle, so that the orbital is expected to
play only a limited role in the low-spin structure of $^{36}$Ar.
One can notice that there are about two holes in the d$_{5/2}$ orbital.
This result implies the structure of $(\text{d}_{5/2})^{-2}
(\text{s}_{1/2}\text{d}_{3/2})^{4.5}(\text{fp})^{4.5}(\text{g}_{9/2})^1$
at the band head of the SD band, which is similar to
the $(\text{s}_{1/2}\text{d}_{3/2})^4(\text{fp})^4$ structure
suggested by the shell model calculation. 
This structure, which has a configuration of two holes in the d$_{5/2}$ orbital
and one particle in the g$_{9/2}$ orbital, is also similar to our result 
for the bandhead structure of the SD band of $^{40}$Ca 
with the g$_{9/2}$ orbital included in the model space.

Turning a focus onto the higher-spin states inside the SD band, 
one can tell that the configuration actually turns out to be closer to the
configuration suggested by the shell-model calculation.
From the Table \ref{tAr2}, our calculation suggests the 
$(\text{d}_{5/2})^{-1}(\text{s}_{1/2}\text{d}_{3/2})^{5}(\text{fp})^{4}$
at $J=16\hbar$. The g$_{9/2}$ orbital seems to play, again, 
no role in this case, so that the state resembles the 4p-4h configuration 
$(\text{s}_{1/2}\text{d}_{3/2})^4(\text{fp})^4$ in the truncated shell-model
diagonalization.
From this analysis, we can say that the basic configuration does not
change much between before and after the backbending. 

Fig.\ref{occupation_gAr} also suggests
that the g$_{9/2}$ orbital is not a key player for the backbending.
Its occupation number is quite low and fairly regular
throughout the whole range of the total angular momentum. 
Instead, the occupation number in the
f$_{7/2}$ orbital suddenly starts to increase in the post-backbending region
($J\agt 7\hbar$) to indicate a rotational alignment in this orbital.
This result is consistent with the interpretation of the backbending
by the PSM \cite{LS01}.
The s$_{1/2}$ orbital also shows the increase, but it is 
more gradual in comparison to the f$_{7/2}$ orbital. The sharp drop of the
hole occupation number in the d$_{5/2}$ orbital 
can be noticed clearly in the figure.
This means that the d$_{5/2}$ orbital starts to be filled by
the deexcitations from the upper shells (about a half particle from the
g$_{9/2}$ orbital and another half from the pf shells). 
This result implies that the deformation becomes less substantial
and the cross-shell excitation is suppressed to some extent.
We will take a closer look at a relation between the deformation ($\beta$)
and the position of the g$_{9/2}$ orbital intruding into the sd shell, 
soon below.

An indication of the decrease in elongation at high spin was reported 
by other calculations such as the shell-model diagonalization \cite{Po04}, 
the cranked Skyrme HF calculation \cite{IMY02}, 
and the cranked Nilsson calculation \cite{SMJ01}. 
\begin{figure*}[tbh]
  \begin{tabular}{cc}
    \includegraphics[width=0.35\textwidth,angle=-90]{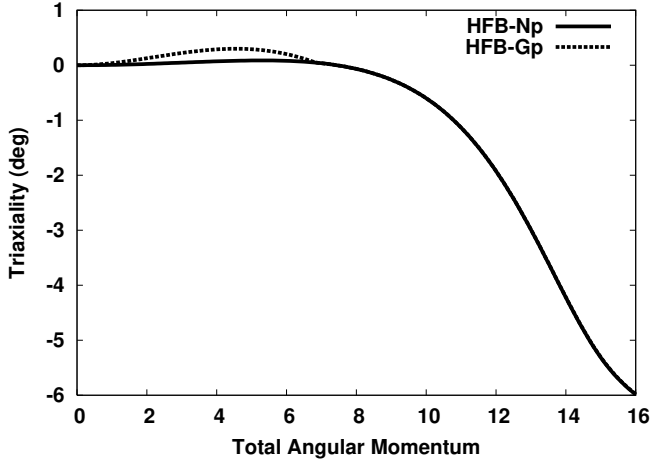}&
    \includegraphics[width=0.35\textwidth,angle=-90]{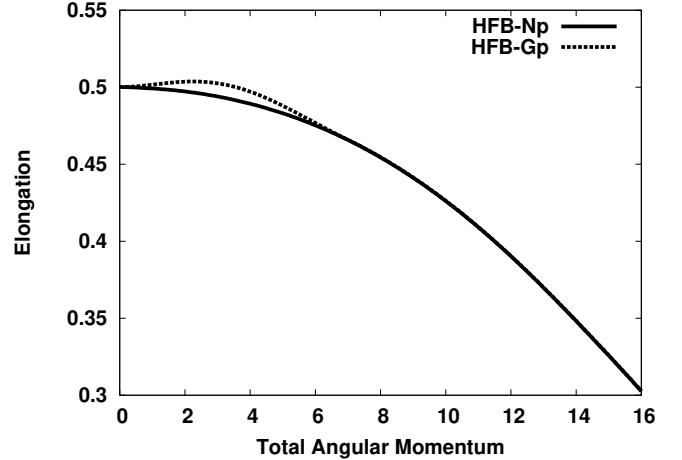}
  \end{tabular}
  \caption{Calculated triaxiality $\gamma$ (left panel) 
    and elongation $\beta$ (right), as functions of the total
    angular momentum  for $^{36}$Ar.}
  \label{Qdef_Ar36}
\end{figure*}
Fig.\ref{Qdef_Ar36} shows the calculated quadrupole deformation
in terms of triaxiality ($\gamma$) and elongation ($\beta$), with our model.
With or without the g$_{9/2}$ orbital in the model space,
there is no much difference in the manner of the shape evolution, 
as already mentioned above.
Before the backbending, the system maintains an axially symmetric shape,
while triaxial deformation starts to grow after the backbending 
(in particular, $J\agt 10\hbar$) although the absolute value of $\gamma$ 
is less than $10^{\circ}$. The small $|\gamma|$ value explains the reason for
the successful descriptions of the PSM calculation \cite{LS01},
as well as the projected GCM calculation \cite{BFH03}.

Based on the above discussions,
it seems that we can conclude that the g$_{9/2}$ orbital 
plays no significant role, unlike the $^{40}$Ca case. 
This can be qualitatively explained from two things.
One is the position of the Fermi levels, which are lower than the one
of the $^{40}$Ca. That is, cross-shell excitations cost energetically more 
in $^{36}$Ar than in $^{40}$Ca.
The other is smaller $\beta$ value of the quadrupole
deformation. From the Nilsson diagram, it can be learned 
that the g$_{9/2}$ orbital can behave like an intruder orbital
when the deformation is as large as $\beta\simeq 0.6$. 
However, when the deformation becomes smaller
such as $\beta\simeq 0.4 - 0.5$, the g$_{9/2}$ orbital remains 
to be high in energy and the mixture of this orbital to the other orbitals 
in the sd shell becomes less likely.
%

If the backbending plot, Fig.\ref{BB_Ar36}, is carefully studied, however,
one can find that there is a little discrepancy
in the high-spin behavior of the curves.
The second backbending seems to happen at $J\simeq 14\hbar$ in the case
without the g$_{9/2}$ orbital (HFB-Np). 
Obviously, this result is inconsistent with the
experimental data, showing no sign of the second backbending.
The case with the g$_{9/2}$ orbital (HFB-Gp) shows the closer results to the
experiment. Apparently from the analyses of the occupation numbers and
the deformation evolution, the g$_{9/2}$ orbital plays no active role
in the SD structure. 
But it seems that these results does not mean that one can exclude the g$_{9/2}$
orbital from the model space. 

Let us examine how this discrepancy happens.
\begin{table}[bt]
  \begin{ruledtabular}
    \begin{tabular}{cccccccc}
      Orbital& d$_{5/2}$ &s$_{1/2}$ &d$_{3/2}$ &f$_{7/2}$ &p$_{3/2}$ &
      f$_{5/2}$ &p$_{1/2}$\\
      \hline
      Proton & 5.38& 1.33& 1.29& 1.95& 0.00& 0.05& 0.00\\
      Neutron& 5.38& 1.33& 1.29& 1.95& 0.00& 0.05& 0.00\\
      \hline
      Total  & 10.76& 2.66& 2.58& 3.90& 0.00& 0.10& 0.00\\
    \end{tabular}
  \end{ruledtabular}
  \caption{Occupation numbers of $^{36}$Ar at $J=16\hbar$ in the case
    with the g$_{9/2}$ orbital excluded from the model space. 
    The subspace (s$_{1/2}$d$_{3/2}$) is occupied by about five ($=5.24$) 
    particles, while the pf-shell is filled with four ($=4.00$) 
    particles.     
    The hole occupation number in the d$_{5/2}$ excluded is 1.24 ($=12-10.76$).}
  \label{tAr3}
\end{table}
First of all, let us compare the occupation numbers in these two cases
(i.e., HFB-Gp and HFB-Np) at $J=16\hbar$.
The case with the g$_{9/2}$ orbital (HFB-Gp) is already shown 
in Table \ref{tAr2},  whereas the case without the g$_{9/2}$ orbital (HFB-Np) 
is presented in Table \ref{tAr3}.
No significant difference can be seen from these two Tables.
Only a tiny difference can be seen 
that the concentration onto the f$_{7/2}$ orbital is slightly higher 
in the latter case than the former.
\begin{figure}[tbh]
  \includegraphics[width=0.35\textwidth,angle=-90]{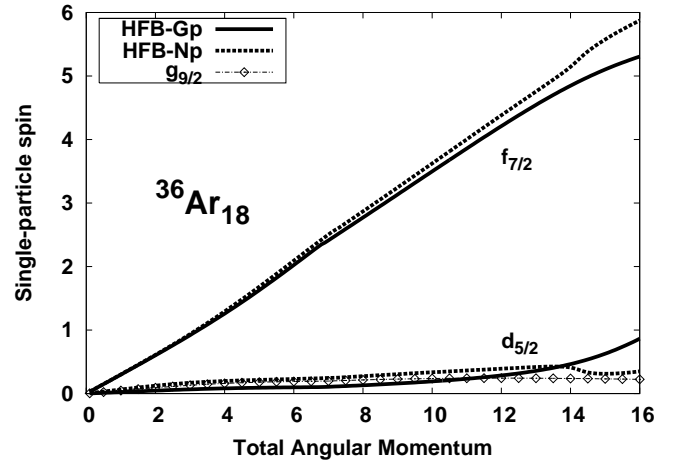}
  \caption{Single-particle spin components along the cranking axis.
  Only the f$_{7/2}$ and d$_{5/2}$ orbitals 
  for protons are plotted to compare the
  two cases (with and without the g$_{9/2}$ in the model space).}
\label{algn_Ar36}
\end{figure}

It is then worth a look at the single-particle spin component
along the cranking axis. There are differences between the two cases,
which are found in the f$_{7/2}$ and d$_{5/2}$ orbitals at $J\agt 14\hbar$.
Their behaviors are plotted in Fig.\ref{algn_Ar36}.
Obviously, the g$_{9/2}$ orbital 
is not involved in the production of angular momentum.
The low occupation number in this orbital implies the consistency 
with the case without the g$_{9/2}$ orbital.
On the other hand, f$_{7/2}$ and d$_{5/2}$ orbitals show slightly
different behaviors in the two cases (HFB-Gp and -Np in the figure, which
correspond to the case with and without the g$_{9/2}$ orbital
 in the model space).
When the g$_{9/2}$ orbital is removed from the model space, 
the alignment in the f$_{7/2}$ orbital is slightly accelerated
 beyond $J=14\hbar$. Whereas, in the other
case, the alignment of the f$_{7/2}$ orbital slows down a little.
The opposite behavior is seen in the d$_{5/2}$ orbitals.
From this analysis, it can be said that the further alignment in the f$_{7/2}$
orbital causes the second backbending when the g$_{9/2}$ orbital is absent.
This backbending should be regarded as an artifact 
and it is caused by the secondary effect due to
the lack of the g$_{9/2}$ orbital, 
which creates subtle differences from the case with the g$_{9/2}$ orbital. 

Therefore, despite the discrepancy in Fig.\ref{BB_Ar36},
our previous conclusion, that is, the g$_{9/2}$ orbital plays no major role 
in the SD band of $^{36}$Ar, still holds
because the qualitative characters of the SD structure does not change
whether the g$_{9/2}$ orbital is considered or not.
It is true, however, that a better description demands
an inclusion of the g$_{9/2}$ orbital in the model space, 
particularly for the backbending plot.

\begin{figure}[tb]
  \includegraphics[width=0.35\textwidth,angle=-90]{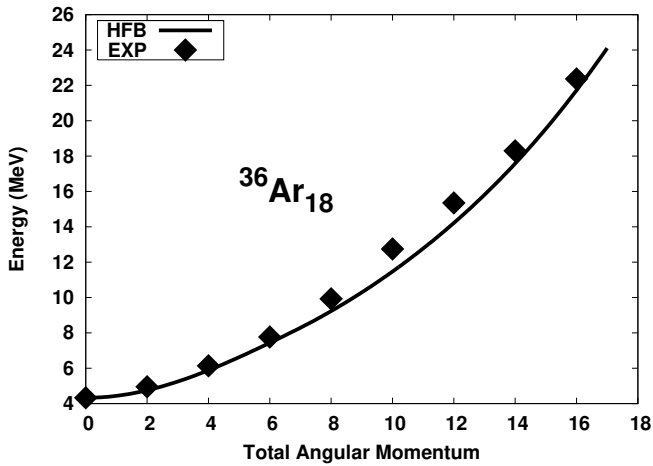}
  \caption{Calculated and observed rotational energies of the SD band 
    in $^{36}$Ar. The g$_{9/2}$ orbital is taken into account in the model space.
    The calculated ground-state energy (at $J=0$) 
    is normalized with the experimental value, $E(J=0)=4.3291$ MeV.}
  \label{SDband_Ar36}
\end{figure}

\begin{figure}[tbh]
  \includegraphics[width=0.35\textwidth,angle=-90]{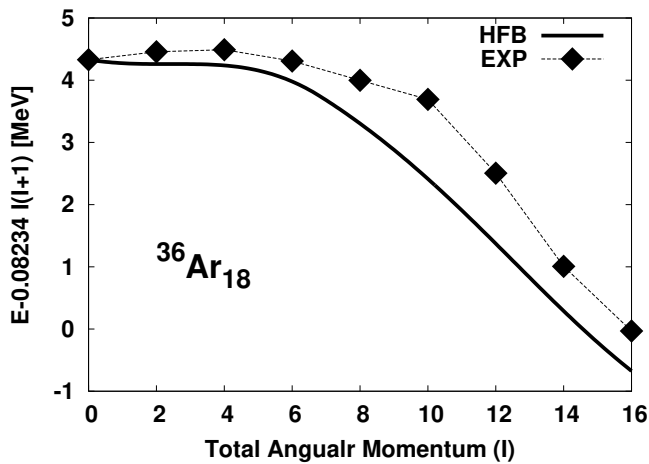}
  \caption{Excitation energies relative to a rigid-rotor energy, 
    $E_{R}=aJ(J+1)$, where $a=0.08234$ (MeV), for $^{36}$Ar.}
  \label{ExEne_Ar36}
\end{figure}

\subsubsection{Pairing correlation and band termination}

Let us now compare the energy spectrum between the calculation
and the experimental data, which are shown in Figs.\ref{SDband_Ar36} 
and \ref{ExEne_Ar36}. Discrepancies are seen in the backbending region 
($8\hbar\alt J\alt 12\hbar$). As we have discussed earlier, these discrepancies
are the result of the pairing collapse (See also Fig.\ref{gap_Ar36}).
But the qualitative behavior is managed to be reproduced.
In particular, the plateau structure at low spin and a sharp slope
at high spin are well reproduced in Fig.\ref{ExEne_Ar36}.
In comparison with the cranked Nilsson calculation in Ref.\cite{SMJ00},
our result shows an improvement as a mean-field approach, which takes
into account the pairing.

\begin{figure*}[htb]
  \begin{tabular}{cc}
    \includegraphics[width=0.35\textwidth,angle=-90]{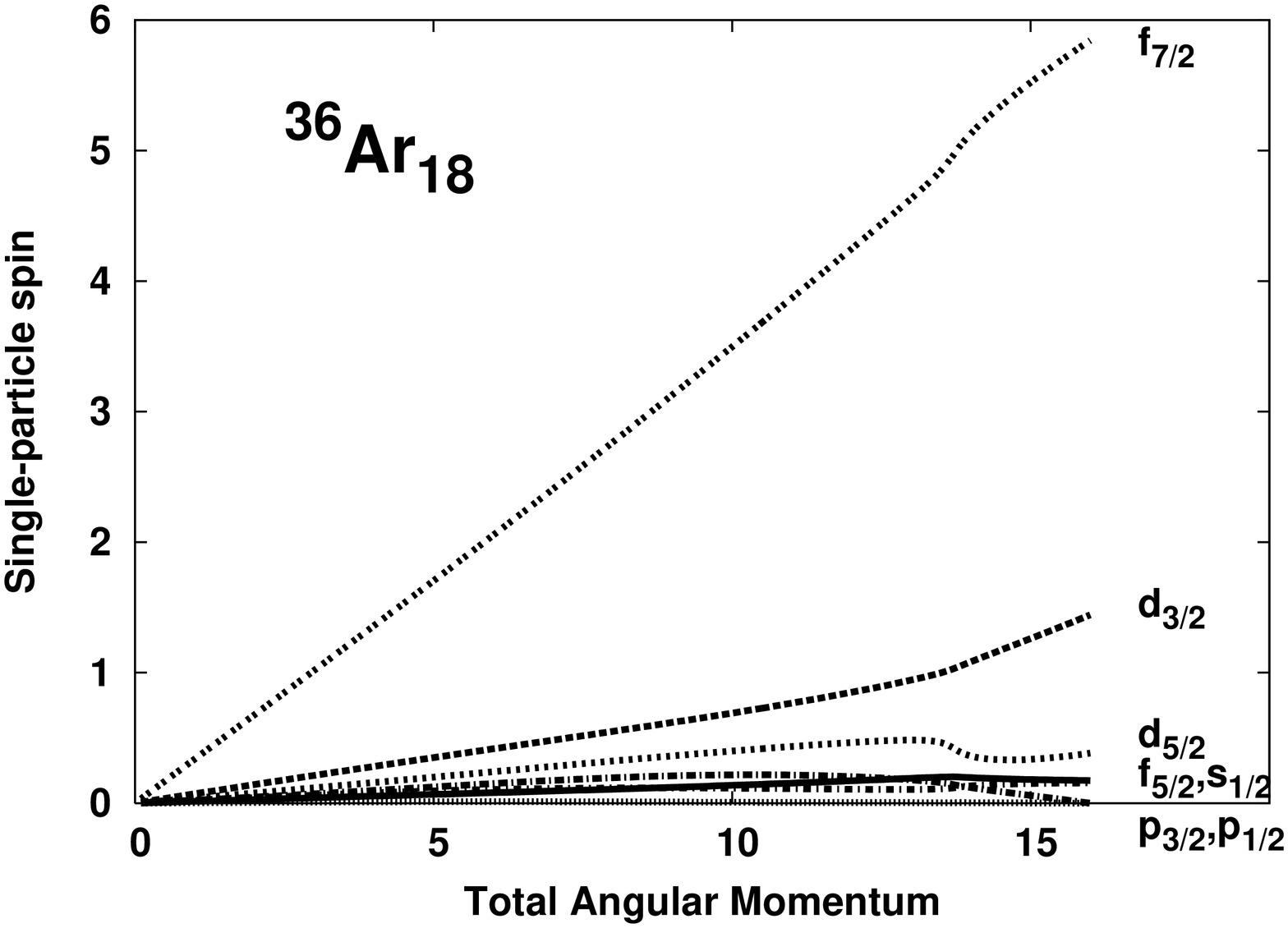}&
    \includegraphics[width=0.35\textwidth,angle=-90]{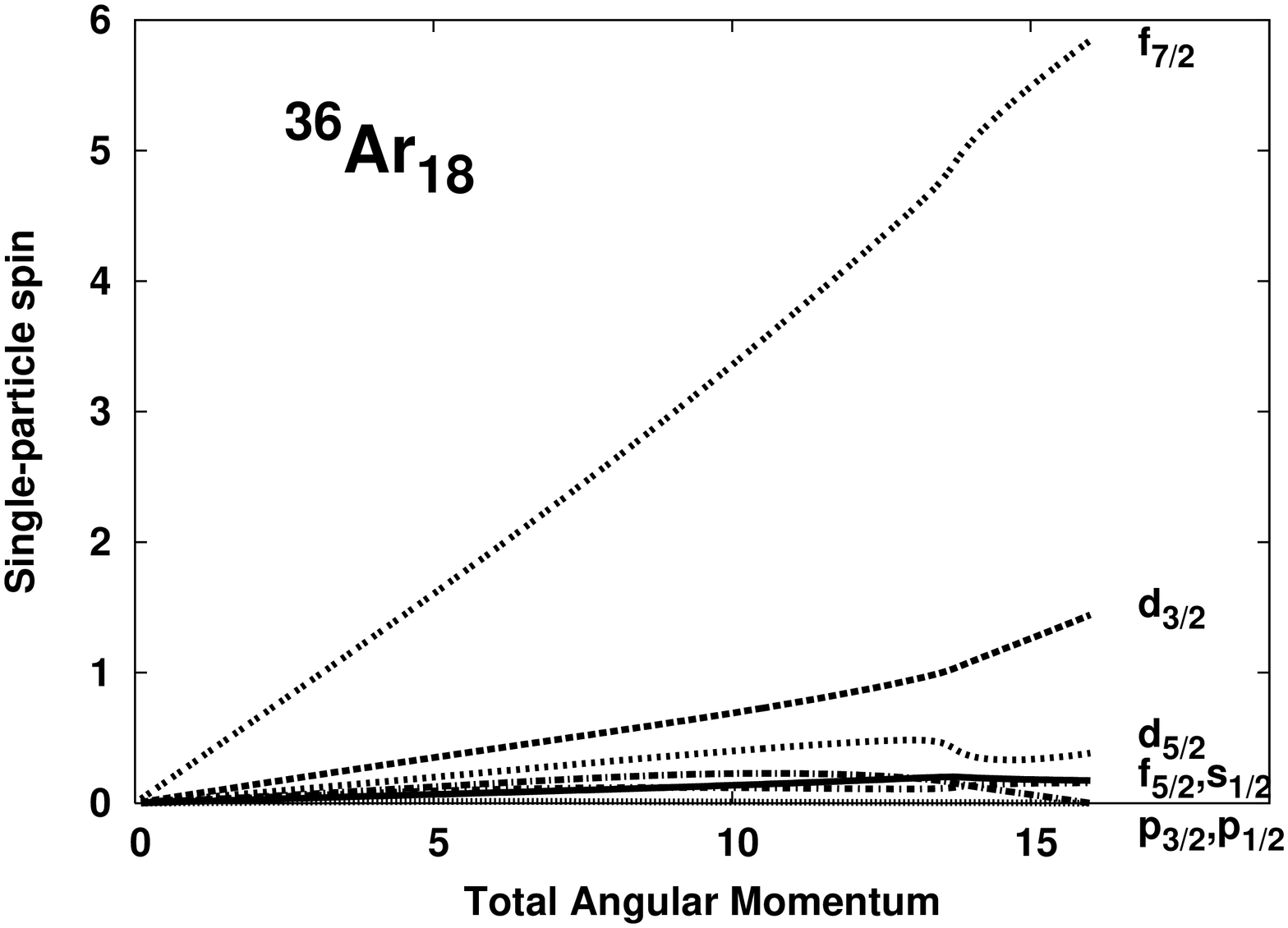}
  \end{tabular}
  \caption{Proton (left) and neutron (right) 
    single-particle spin components along the cranking axis
    for $^{38}$Ar. 
    The g$_{9/2}$ orbital is included in the model space.}
  \label{algn_nnp}
\end{figure*}

\begin{figure*}[tbh]
  \begin{tabular}{cc}
    \includegraphics[width=0.35\textwidth,angle=-90]{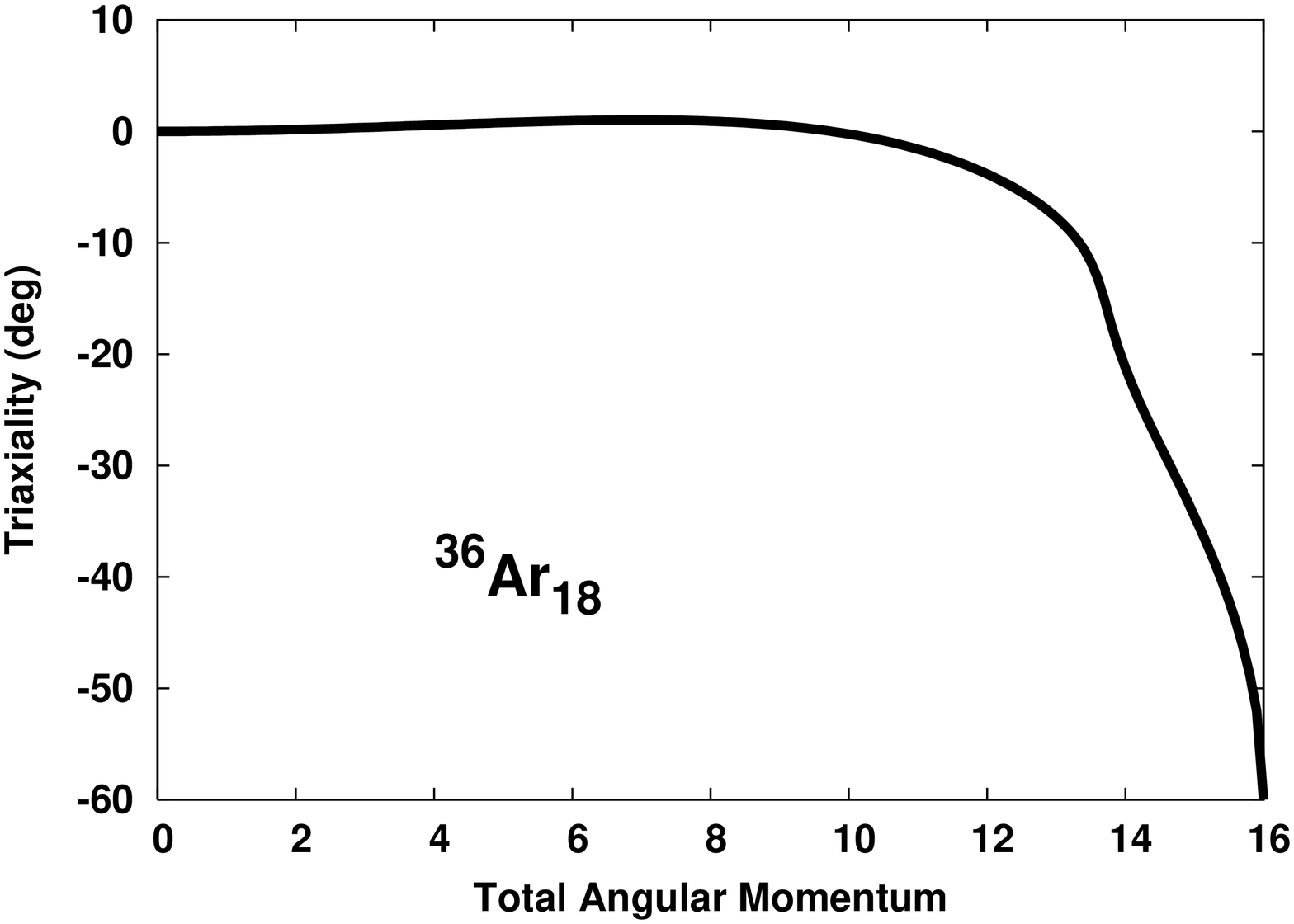}&
    \includegraphics[width=0.35\textwidth,angle=-90]{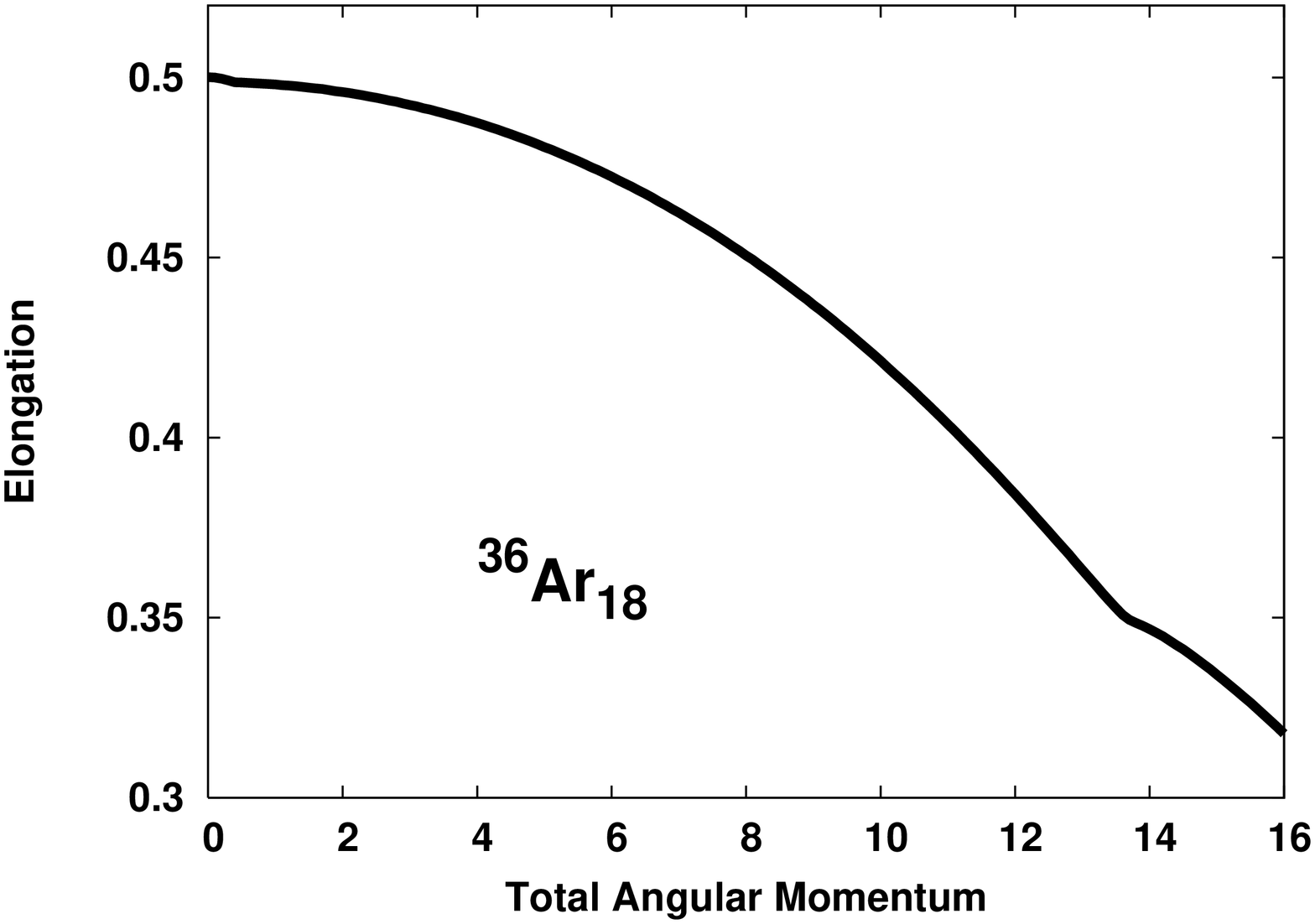}
  \end{tabular}
  \caption{Calculated triaxiality $\gamma$ (left panel) 
    and elongation $\beta$ (right), as functions of the total
    angular momentum  for $^{36}$Ar, 
    when the pairing correlation is switched off.}
  \label{Qdef_Ar36_np}
\end{figure*}

Without the pairing, the cranked Nilsson calculation
suggests an occurrence of the band-termination phenomenon
for the SD band of $^{36}$Ar \cite{SMJ01}.
The typical feature of the band termination appears as a shape change
into oblate deformation.
Earlier in the present study,
we have learned that the triaxial deformation is suppressed 
with a proper treatment of the pairing (Fig.\ref{Qdef_Ar36}).
It is interesting to study if the band termination and the
associated oblate deformation can emerge in the absence of the pairing
in our model.

By choosing the initial pairing-gap parameters to be very weak, that is,
$(\Delta_{\text{p}}^0,\Delta_{\text{n}}^0)=(0.15,0.15)$ MeV,
the effect of the pairing correlation is examined in the following calculation.
(The g$_{9/2}$ orbital is included in this analysis.)
This choice of the initial parameters causes
the breakdown of the pairing as early as at $J\simeq 1.5\hbar$.
Effectively, the calculation turns to be the HF calculation beyond
this total angular momentum, which can be comparable with the
cranked Nilsson calculation without the pairing.
The corresponding cranking calculation gives 
regular solutions until $J=16\hbar$, the band termination point.
But beyond the termination point, the solution becomes irregular
and shows unphysical behavior, so that we ignore the calculations beyond
the band termination point in this analysis.

In the band termination spin ($J=16\hbar$), the occupation numbers
of the f$_{7/2}$ orbital 
are calculated to be 1.94 each for neutrons and protons.
Therefore, the net particles occupying the f$_{7/2}$ orbital 
are four (i.e., $=2+2$), 
and the result is consistent with the shell model configuration.
Two particles occupying the f$_{7/2}$ orbital can generate
the maximum angular momentum of $J=6\hbar$, which
corresponds to the band termination. 

In Fig.\ref{algn_nnp}, the calculated single-particle alignments are
plotted for protons and neutrons.
Our calculation (without the paring)
reproduces the above situation: the single-particle angular momentum carried
by the f$_{7/2}$ orbital is almost 12$\hbar$, consisting of
$6\hbar$ for protons and $6\hbar$ for neutrons.

In Fig.\ref{Qdef_Ar36_np}, the calculated quadrupole deformation,
that is, triaxiality ($\gamma$) and elongation ($\beta$), are plotted.
The profile for the elongation does not change from the previous case with
the pairing correlation. However, the behavior of the triaxial evolution
is different, especially beyond $J=12\hbar$. 
At the band termination point, the value of $\gamma$ reaches $-60^{\circ}$,
indicating the non-collective oblate shape.
From this analysis relying on the self-consistent cranking calculation, 
it is confirmed that 
the band termination phenomenon happens when the pairing correlation
is absent from the system (or very weak).

\subsubsection{Summary for $^{36}$Ar}
As a partial summary for $^{36}$Ar,
we can conclude that the previously proposed structure,
$(\text{s}_{1/2}\text{d}_{3/2})^{4}(\text{fp})^4$, 
seems to be a good approximation for the structure of $^{36}$Ar,
according to our results. It is suggested that
backbending in the SD band of $^{36}$Ar is
caused by the alignment in the f$_{7/2}$ orbitals, as concluded by the
PSM calculation \cite{LS01}. The g$_{9/2}$ orbital does not play
any significant role. 
From the successful description of the SD band at low spin and
the failure in the backbending region,
it was demonstrated that the pairing correlation is 
very important to describe the structure of the SD band.
Triaxial deformation starts to occur at high spin, but the degree of
triaxiality is not so substantial that the SD states
are well described as an axially symmetric nuclear many-body system.

\section{Conclusions}
The self-consistent cranking calculation based on the HFB method
was applied to the superdeformed bands of two $N=Z$ nuclei, $^{40}$Ca
and $^{36}$Ar. Our microscopic calculations with the P$+$Q$\cdot$Q interaction 
can manage to give good qualitative explanations (occasionally quantitatively) 
to the energy spectrum, rotational alignment, and backbending phenomenon 
of these nuclear systems. 

Special attentions were paid to the roles of (1) the d$_{5/2}$ orbital, which
was removed from the sd-pf model space in the shell-model diagonalizations;
and (2) the g$_{9/2}$ orbital, which belongs to a higher shell ($N=4$) than
the sd-pf shell. The effect of the pairing correlation was also investigated
in connection to the evolution of triaxial deformation and
the band termination phenomenon.

Inside the framework of our model, 
it was found that the truncation of the d$_{5/2}$ orbital can be justified
as far as lower-spin states are considered. Whereas,
high-spin states are found to be produced due to a gradual excitation
from the d$_{5/2}$ orbital to the upper sd shell. 
However, in either case of $^{40}$Ca and $^{36}$Ar, an inclusion of the
d$_{5/2}$ orbital does not affect the nuclear structure of the SD states 
very much.

On the contrary, the g$_{9/2}$ orbital was found to change the nuclear
structure drastically for $^{40}$Ca: backbending may happen 
at $J\simeq 20\hbar$. However, the orbital plays no significant role
for $^{36}$Ar. These differences come from the location of the Fermi levels 
and the deformation ($\beta$),
which is $\simeq 0.6$ for $^{40}$Ca while $\simeq 0.4$ for $^{36}$Ar.
This difference influences the position of the g$_{9/2}$ orbital
as an ``intruder orbital'' into the sd shell, 
in terms of the deformed Nilsson model.

The pairing correlation was found to be important to produce
a proper energy spectrum and tend to act as a suppressor of 
triaxial deformation. Without the pairing, triaxial deformation would be
enhanced and the non-collective oblate shape ($\gamma=-60^{\circ}$)
would ultimately emerge at high spin. However,  in our model taking the
pairing correlation into account, it is observed that triaxiality is suppressed 
to $|\gamma|\alt 10^{\circ}$. In this sense, the SD states of both
$^{40}$Ca and $^{36}$Ar are nearly axial symmetric in our model,
which can justify other calculations assuming the axial symmetry.

Despite the success for the qualitative explanations,
a problem was recognized in relation to the pairing collapse at high spin.
An improvement is surely necessary for more accurate and quantitative
descriptions of the high-spin structure, especially around the backbending
region. 

Nevertheless, through this work, 
the P$+$Q$\cdot$Q model based on the cranked HFB approach
was demonstrated to be a practical and effective model to describe
high-spin nuclear structure showing superdeformation.

\begin{acknowledgments}
M.O. appreciates useful discussions with 
S. Williams, E. Ideguchi, T. Shizuma, N. Onishi and N. Timofeyuk.
The author thanks P. M. Walker for his careful reading of the manuscript.
This work is supported by STFC/EPSRC
with an advanced research fellowship GR/R75557/01 
as well as a first grant EP/C520521/1.
\end{acknowledgments}

\end{document}